\documentclass[apj]{emulateapj}
\newcommand{\degree}{\hbox{$^\circ$$$}}

\shorttitle{Numerical CMD analysis of SDSS data}
\shortauthors{de Jong et al.}

\begin{document}

\title{Numerical Color-Magnitude Diagram Analysis of SDSS Data and
  Application to the New Milky Way satellites}

\author{J. T. A. de Jong\altaffilmark{1}, H.-W. Rix\altaffilmark{1},
N. F. Martin\altaffilmark{1}, D. B. Zucker\altaffilmark{2}, 
A. E. Dolphin\altaffilmark{3}, E. F. Bell\altaffilmark{1},\\
V. Belokurov\altaffilmark{2}, N. W. Evans\altaffilmark{2}}
\email{dejong@mpia.de}

\altaffiltext{1}{Max-Planck-Institut f\"{u}r Astronomie,
  K\"{o}nigstuhl 17, D-69117 Heidelberg, Germany}
\altaffiltext{2}{Institute of Astronomy, University of Cambridge,
  Madingley Road, Cambridge CB3 0HA, United Kingdom}
\altaffiltext{3}{Steward Observatory, University of Arizona, 933 N. Cherry
  Ave, Tucson, AZ, 85721, United States}

\begin{abstract}
We have tested the application to Sloan Digital Sky Survey data of the
software package MATCH, which fits color-magnitude diagrams (CMDs) to
estimate stellar population parameters and distances. These tests on a
set of six globular clusters show that these techniques recover their
known properties. New ways of using the CMD-fitting software enable us
to deal with an extended distribution of stars along the
line-of-sight, to constrain the overall properties of sparsely
populated objects, and to detect the presence of stellar overdensities
in wide-area surveys.  We then also apply MATCH to CMDs for twelve
recently discovered Milky Way satellites to derive in a uniform
fashion their distances, ages and metallicities. While the majority of
them appear consistent with a single stellar population, CVn I, UMa
II, and Leo T exhibit (from SDSS data alone) a more complex history
with multiple epochs of star formation.
\end{abstract}


\section{Introduction}

Studying the properties of resolved stellar populations through
color-magnitude diagrams (CMDs) can provide important constraints on stellar
evolution and on the formation and evolution of stellar systems. For
example, the apparent magnitude of the horizontal branch (HB) stars in
a population, or the magnitude and color of the main sequence turn-off
(MSTO), constrain or allow determination of the distance to and age
and metallicity of the population. The most detailed and most
objectively quantifiable information can be obtained by a numerical
CMD fitting analysis \citep[see
e.g.][]{gallart96, tolstoy96, aparicio97, dolphin97, holtzman99,
olsen99, hernandez00, harris01}.

Over the last seven years, the Sloan Digital Sky Survey
\citep[SDSS,][]{york00,dr5}, has performed a large wide-field
photometric and spectroscopic survey in the northern Galactic
cap. Because SDSS, unlike previous all-sky surveys, is a CCD-based
survey, the photometric depth, accuracy, and homogeneity achieved is
unprecedented over such a large area. SDSS data have proven to be of
great value for the study of the Galactic stellar halo and nearby
stellar systems with resolved populations
\citep[e.g.][]{newberg02,fieldofstreams,juric05,bell07}. They have
also led to the discovery of a host of previously unknown, faint Local
Group members \citep{UMaI,Wil1,CVnI,UMaII,Boo,5pack,LeoT,BooII}.

The study of dwarf galaxies and the Milky Way halo can provide
important clues to the formation and assembly of the Local Group of
galaxies and its individual constituents, as well as test cosmological
theories. In the cold dark matter (CDM) paradigm, an abundance of
subhaloes should be present around Milky Way type galaxies. The
observed number of satellites is, however, one or two orders of
magnitudes lower than what CDM theory predicts
\citep{klyp99,moore99}. The recently discovered satellites partially
alleviate this discrepancy, but if the CDM predictions are accurate,
many more new satellite galaxies are to be discovered. Another
prediction of CDM theories is that galaxies are formed hierarchically,
i.e. built up by accreting smaller systems
\citep[e.g.][]{searlezinn,bullockjohnston}. Among the population of
satellite galaxies surrounding the MW, some should have been formed
within the MW halo while others should have a different
origin. Studying the stellar populations, masses and abundance
patterns of these satellites can provide important constraints for
galaxy formation and CDM theory.

Our goals for this paper are two-fold. First, we adapt the CMD-fitting
software MATCH \citep{match} for use on SDSS data, making use of the
theoretical isochrones provided by \cite{girardi02}, and describe
several methods of applying this software to the data. A set of
globular clusters taken from the SDSS data is used to test both the
algorithmic approach and the theoretical isochrones. Second, we apply
MATCH to the SDSS data for the new Milky Way satellites (that were
recently discovered using the same SDSS data), to provide a uniform
analysis of these objects and constrain their population parameters.

In Section \ref{sec:methods} we describe the data and the different
CMD-fitting methods. The tests on globular clusters are done in
Section \ref{sec:gctests} and in Section \ref{sec:newsats} we
analyze the properties of the new Galactic satellites. We conclude
and summarize our results in Section \ref{sec:summary}.

\newpage

\section{Data and Methods}
\label{sec:methods}

\subsection{Sloan Digital Sky Survey data}

The SDSS is a large photometric and
spectroscopic survey, primarily covering the North Galactic Cap. As of
the Data Release 5 \citep[DR5,][]{dr5}, SDSS photometry
covers almost a quarter of the sky in five passbands \cite[$u$, $g$,
$r$, $i$, and $z$;][]{Gu98,Gu06,Ho01}.  For the work presented in
this paper we use photometry in the two most sensitive bands, namely $g$
and $r$.  A fully automated data reduction pipeline produces accurate
astrometric and photometric measurements for all detected objects
\citep{Lu99,St02,Sm02,Pi03,Iv04,Tu06}. Photometric accuracy is on the order
of 2\% down to $g\simeq$22.5 and $r\simeq$22 \citep{Iv04,sesar07}, and
comparison with deeper surveys has shown that the data are complete to
$r\simeq$22.  We use stellar data extracted from the SDSS archive with
artifacts removed and we apply the extinction corrections from
\cite{sfd} provided by DR5.

\begin{figure}
\epsscale{1.0}
\plotone{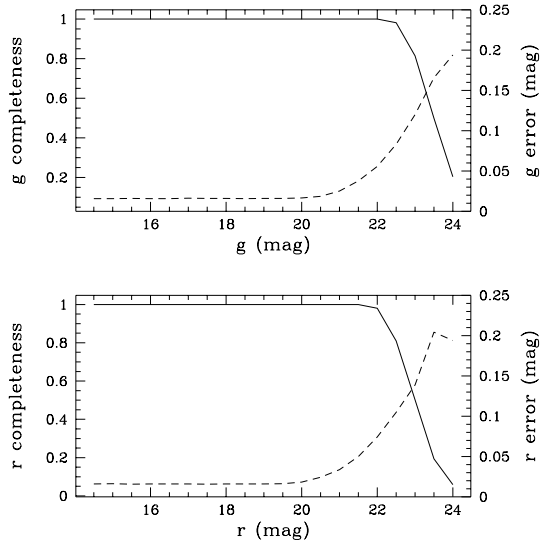}
\caption{ Completeness and photometric errors of the artificial
star files adopted for the SDSS DR5 data and used for
creating realistic stellar population models. {\it Upper panel:}
$g$-band completeness fraction (solid line) and average photometric
error (dashed line) as function of $g$-band magnitude. {\it Lower
panel:} the same for $r$-band.  }
\label{fig:fakedetails}
\end{figure}

In order for the CMD fitting with MATCH to work, a model describing
the completeness and the photometric accuracy as a function of
magnitude is needed. Completeness has been checked for
$r$-band\footnote{See
http://www.sdss.org/dr5/products/general/completeness.html} by
comparing detection of stars and galaxies by SDSS with the deeper
COMBO-17 survey\footnote{
http://www.mpia.de/COMBO/combo\_index.html}. For stellar sources the
completeness is 100\% to $r$=22 and then drops to 0\% at r=23.5. Since
the stellar classification of COMBO-17 starts to become incomplete at
approximately one magnitude fainter than in SDSS, this completeness is
not extremely accurate, but amply sufficient for our purposes. The
photometric errors given by the automated reduction pipeline have been
found to be very accurate \citep{Iv04}. Comparison of the photometric
accuracy in $g$ and $r$ as a function of apparent magnitude shows
similar behavior, but in $g$ the rise in the uncertainties occurs some
0.5 magnitudes fainter.  For the $g$-band completeness we assume
similar behavior as in $r$, but also shifted half a magnitude fainter.
Using this information we create artificial star test files that are
given as input to our CMD fitting code. From this, the code creates
the photometric accuracy and completeness model with which the
theoretical population models are then convolved.  Figure
\ref{fig:fakedetails} shows the completeness and average photometric
errors of our artificial star files for the $g$- and $r$-band
separately.  Even though differences in, for example, seeing cause
differences in photometric accuracy and completeness between different
parts of the sky, we use this same average model in the remainder of
this paper.

\subsection{Color-Magnitude Diagram fitting}

Several software packages have been developed for quantitative and
algorithmic CMD (or Hess diagram) fitting \citep[e.g.][]{gallart96,
tolstoy96, aparicio97, dolphin97, holtzman99, olsen99, hernandez00,
harris01}. The strength of these methods lies in the fact that all
photometric information from imaging in two bands can be used
simultaneously in a robust and numerically well-defined way.  The
software package we use is MATCH, which was written and successfully
used for the CMD analysis of globular clusters and dwarf galaxies
\citep{match}. The software works by converting an observed CMD into a
Hess diagram \citep[a two-dimensional histogram of stellar density as
a function of color and magnitude;][]{hess24} and comparing it with
synthetic Hess diagrams of model populations. To create these
synthetic Hess diagrams, theoretical isochrones are convolved with a
model of the photometric accuracy and completeness. The use of Hess
diagrams enables a pixel-by-pixel comparison of observations to models
and a maximum-likelihood technique is used to find the best-fitting
single model, or best linear combination of models. To account for
contamination by fore- and background stars, a ``control field'' CMD
needs to be provided. MATCH will then use the control field Hess
diagram as an extra `model population' that can be scaled up or
down. The resulting output will be the best-fitting linear combination
of population models plus the control field.

To construct star formation histories (SFHs) and age-metallicity
relations (AMRs), a set of theoretical isochrones is needed that spans
a sufficient range of ages, masses and metallicities. When MATCH was
written, the only published set of isochrones with full age and
metallicity coverage available was provided by \cite{girardi02}. Since
we are specifically interested in the application to SDSS data, we
need isochrones in the SDSS photometric system. The only set available
to date is one from \cite{girardi04} and has the same parameter space
coverage. Some worries exist about the accuracy of the isochrone
calibration \citep[e.g.][]{clem06}, which was limited to three systems
available in the first data release (DR1) of SDSS. This unavoidably
adds an extra source of uncertainty to all CMD fitting results. One of
the main motivations for the various tests that are described in this
paper is to see whether reasonable results can be obtained using the
current \cite{girardi04} isochrones.

\subsection{Modes of CMD-fitting}
\label{sec:modes}

We use MATCH for slightly different purposes than originally described
by \cite{match}; the new ``modes'' are described briefly below.

\subsubsection{Star formation histories and metallicity evolution}
\label{subsec:sfhamr}
MATCH was originally developed to solve for the SFH and AMR of a
stellar system for which all stars can be assumed to be at the same
distance, such as globular clusters and dwarf galaxies. In this
original mode of MATCH the distance and the foreground extinction to
all stars are fixed a priori for each fit.  From a set of age bins and
metallicity bins, the combination that best fits the observed CMD is
found, resulting in an estimate of the SFH and AMR. By doing the same
fit with different values of the foreground extinction and distance,
these two latter parameters can also be determined by comparing the
goodness-of-fit of the individual fits. In principle, parameters like
the steepness of the IMF and the binary fraction can be constrained in
the same fashion, although in most cases the data quality and the
degeneracies between different parameters will prevent the
determination of all of these parameters simultaneously. Analyzing the
parameter range of acceptable fits also gives robust error estimates
for the SFH and AMR; this approach we refer to as MATCH's ``standard
mode''.

\subsubsection{Distance solutions}
\label{subsec:distsol}
To apply CMD fitting techniques to nearby and extended stellar systems
(for example substructure within the Milky Way as in \cite{dejong07}),
the assumption that all stars are practically equidistant need not
hold. To fit populations at differing distances simultaneously we have
implemented a new way in which to run MATCH that keeps the same number
of free parameters, or parameter dimensions, to be explored.  While in
the standard mode the distance is fixed and age and metallicity are
independent variables, here the distance becomes a free parameter, but
age and metallicity are being limited via a specified age metallicity
relation (AMR). For any given AMR there is only one ``population
parameter'', e.g. age with a certain metallicity linked to each age
bin. Of course, different age-metallicity relations can be explored,
potentially covering a wide range of possibilities.  It seems sensible
to choose an AMR that is appropriate for the studied stellar
population on astrophysical grounds, such as the expected or known
metallicity evolution.  For example, for the study of structures in
the Galactic thin disk one would consider a different AMR than for the
study of a metal-poor dwarf galaxy. However, within this new mode of
running MATCH the choice of AMR is completely flexible and a flat AMR
(no dependence of metallicity on age) or even an AMR where metallicity
decreases with time is possible.

\subsubsection{Single Component fitting}
\label{subsec:scfit}
When studying low-luminosity or low surface brightness contrast
systems, such as some of the recently discovered dSphs, or systems at
low Galactic latitudes, for example substructures in the Milky Way,
the contamination of the CMDs of these systems by fore- and background
stars will be very high. If the ratio of `member' stars to field stars
becomes low, the S/N of the CMD features to be fit consequently
becomes very low.  In such cases, all free parameters used to
constrain distance, age, and metallicity in the MATCH mode of Sections
\ref{subsec:sfhamr} and \ref{subsec:distsol} can no longer be
constrained by the data alone. To test a more restricted set of
hypotheses in the case of low S/N systems we have developped a single
component (SC) fitting mode for MATCH. In this mode, template stellar
population models are created that consist of a single component with
given, fixed distance, foreground extinction, age range and
metallicity range\footnote{While we use 'single components' it is
important to note that these are not single stellar populations (SPP)
with only a single age, distance, and chemical composition}. Such a SC
template is fit to the data together with a control field. Since the
template consists of a single component, the only free parameters
during the fit are the absolute scaling of (i.e. the number of stars
in) the template and the control field. For each individual SC
(i.e. for each combination of distance, foreground extinction, age
range and metallicity range), a value of the goodness-of-fit is
obtained. Comparing the goodness-of-fit values for all the SC's we can
then find the best-fitting templates and thus constrain the overall
properties of the system.

The quality of each fit is expressed by the goodness-of-fit parameter
Q, determined using the maximum-likelihood technique described in
\cite{match}. This statistic is meant to measure how likely it is that
the observed distribution of stars would result from a random drawing
from the fitted model. In the SC fitting scheme we know that our fits
will not be perfect since an SC model will not be a perfect
representation of a real stellar population, but we are primarily
interested in a likelihood-ratio test for a range of SC models. The
expected variance in the parameter $Q$ for a certain CMD can be
calculated from Poisson statistics and also obtained from Monte Carlo
simulations using random drawings from the models. All fits with a
value of $Q$ less than 1 $\sigma$ away from that of the best-fit SC
model can be considered statistically acceptable, under the assumption
that the best fit is a reasonable representation of the data. Taking
all these fits together, error estimates on the fit parameters can be
determined from the encountered spread of the parameter values.

This approach of using relative values of $Q$ should only be used in
cases where the ``best fit'' is a reasonable approximation of the
data. In cases where there are several very different prominent
stellar populations this will not be the case and results will break
down. Studying different parts of the CMD separately can help to
circumvent such problems, as is done in \cite{dejong07} and also in
Section \ref{subsec:nsscfits} of this paper for the dwarf galaxy Leo
T.

\subsubsection{Over-density detection}
\label{subsec:detect}
The same CMD fitting techniques can also be used to detect or verify,
faint stellar over-densities in large datasets like that of the SDSS
survey. In case of an overdensity at a specific distance, the stars
will not only cluster around a certain spatial position, but also
along a certain locus in the CMD.  The method we have developped works
by fitting the CMD of a small target patch of sky both with a control
CMD obtained from a large area around the target patch and with the
same control field plus a simple stellar population model. If no
over-density is present, the fit with the control CMD should be good
and no significant improvement in fit quality will result by including
an extra stellar population model. However, if there is an
over-density present and the parameters of the population model, such
as distance, age and metallicity, are close enough to those of the
over-density, the fit including the model will result in a significant
improvement in the fit quality. By doing this for a range of models
the presence of a stellar over-density can be assessed by looking at
the difference in the fit quality parameter, $\Delta Q$, between the
control CMD fit and the control CMD plus model fit. Moreover, by
comparing the values of $\Delta Q$ for different models, a rough
estimate of the properties of the overdensity can be obtained at the
same time.

An important prerequisite for this method is of course that the
over-density should contribute a larger fraction of the total number
of stars to the target CMD than to the control CMD, and ideally should
be contained completely within the target patch. This means that
over-densities will be detected with the highest S/N if their size
matches the size of the target patch.

\subsection{Fits to artificial data}
\label{subsec:fakegc}

\thispagestyle{empty}
\begin{deluxetable}{cccc}
\tablecaption{Artificial CMD properties}
\tablewidth{0pt}
\tablehead{ \colhead{CMD} & \colhead{Ages (Gyr)} & \colhead{[Fe/H] (dex)} & \colhead{m-M (mag)} }
\startdata
1 & 10--13 & $-$1.5 & 14.5 \\
2 & 10--13 & $-$1.7 & 14.5 \\
~ & 2--2.5 & $-$1.3 & 14.5 \\
3 & 10--13 & $-$1.5 & 20.5 \\
4 & 10--13 & $-$1.7 & 20.5 \\
~ & 2--2.5 & $-$1.3 & 20.5 \\
\enddata
\label{tab:fakegcprops}
\end{deluxetable}

\begin{figure}[t]
\epsscale{1.1}
\plotone{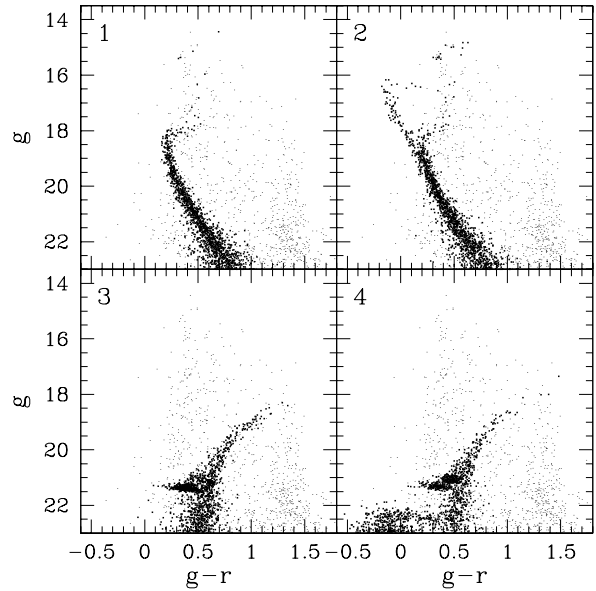}
\caption{Artificial color-magnitude diagrams used to test our
  CMD-fitting methods. Stars belonging to the artificial stellar
  populations listed in Table \ref{tab:fakegcprops} are plotted
  thicker than the field stars taken from SDSS data.
}
\label{fig:fakecmds}
\end{figure}

Before turning to real data it is useful to get a feel for the
degree of precision that we can expect CMD fits to achieve.  To do
this, we create a few realistic artificial CMDs and apply SFH fitting,
SC fitting and distance fitting methods to these to try and recover
their properties.  Using the theoretical isochrones from
\cite{girardi04}, a Salpeter initial mass function \citep{salpeter},
and our model for the photometric errors and completeness of the SDSS
data, we generate artificial stellar populations, to which field stars
taken from the SDSS data are added. In total we create four CMDs,
shown in Figure \ref{fig:fakecmds}. The number of stars in each CMD is
chosen to be the same, to make for a fair comparison.  CMDs 1 and 3
contain a simple, single burst stellar population, located at
different distances, so that either the MS and MSTO or the RGB and HB
are prominent features. Artificial CMDs 2 and 4 contain two different
age and metallicity populations, again with the CMDs corresponding to
different distances. The detailed parameters of the artificial
populations are listed in Table \ref{tab:fakegcprops}.  From the SDSS
archives all stellar sources in a 5\degree$\times$5\degree~ patch of
sky close to the globular cluster M13 were extracted and a small
fraction used to create the field star contamination in these
CMDs. The remaining sources will be used as a ``control field''
population when fitting these background contaminated CMDs.

\begin{figure*}[ht]
\epsscale{1.0}
\plotone{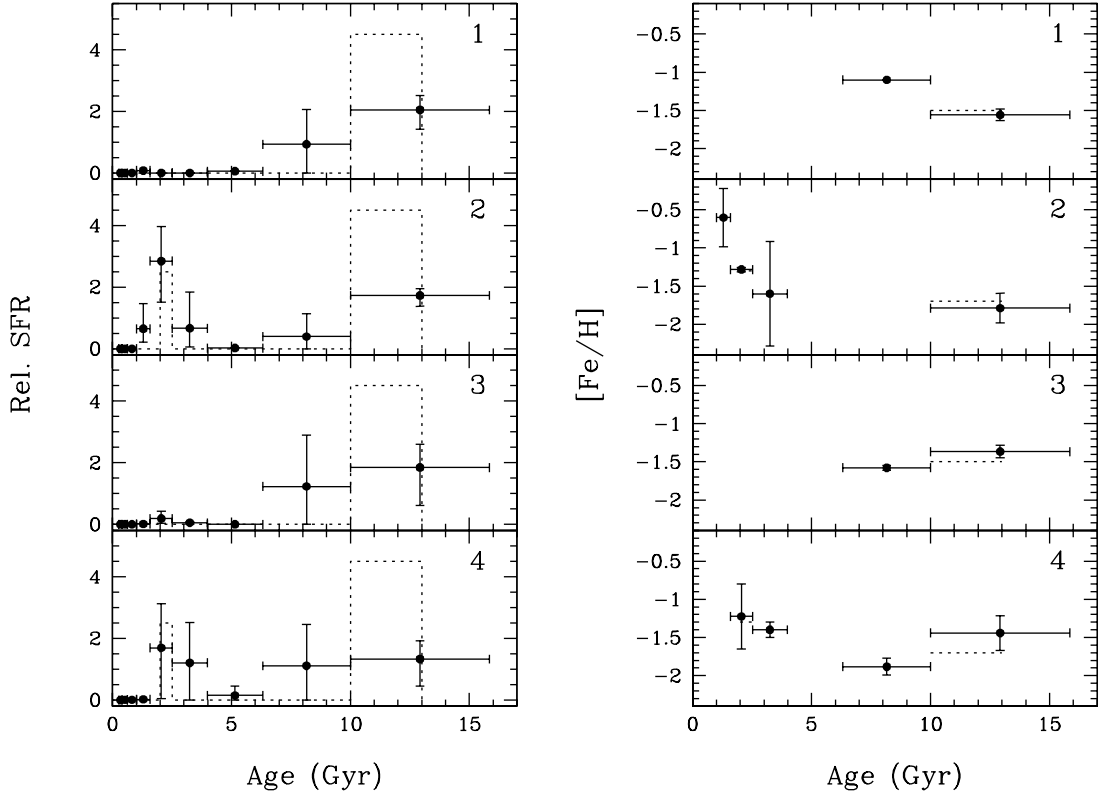}
\caption{Star formation histories for the artificial CMDs from the
  ``standard mode'' of CMD-fitting. In the columns on the left the SFR
  relative to the average SFR over the total age range is plotted, and
  in the columns on the right the SFR-weighted metallicity for the
  bins with a significant SFR. Horizontal error bars indicate the
  width of the age bins. The vertical error bars on the SFR give the
  full range of SFRs found in the fits, while the vertical error bars
  on the metallicities give the standard deviation in the
  metallicities of the fits. Dashed lines indicate the input SFHs and
  metallicities for the artificial CMDs. }
\label{fig:fakesfhamr}
\end{figure*}

We recover the SFHs of the artificial CMDs by running MATCH with nine
age bins and twelve metallicity bins. The age bins are defined in log
$t$ because the isochrones are spaced more evenly in log $t$ than in
$t$, with the oldest running from 10.0 to 10.2 (10 Gyr to 15.8 Gyr)
and the youngest from 8.6 to 8.8 (400 Myr to 630 Myr). The metallicity
range of [Fe/H]$=-$2.4 to 0.0 is covered by twelve bins of 0.2 dex
width. Thus, the artificial CMDs are fit with 108 model CMDs plus the
``control field'' stars, for a total of 109 partial CMDs. The best
fitting linear combination of model CMDs will form the resulting SFH
and metallicity evolution. To get a sense of the uncertainties on the
SFH, each CMD is fit thrice, each time with a slightly different
assumed distance. The distance moduli used are 14.3, 14.4 and 14.5
mag.  The resulting SFHs and metallicities are presented in Figure
\ref{fig:fakesfhamr} as the relative star formation rate (SFR) and as
[Fe/H] as function of the age of the stars.

\begin{figure*}[ht]
\epsscale{1.0}
\plotone{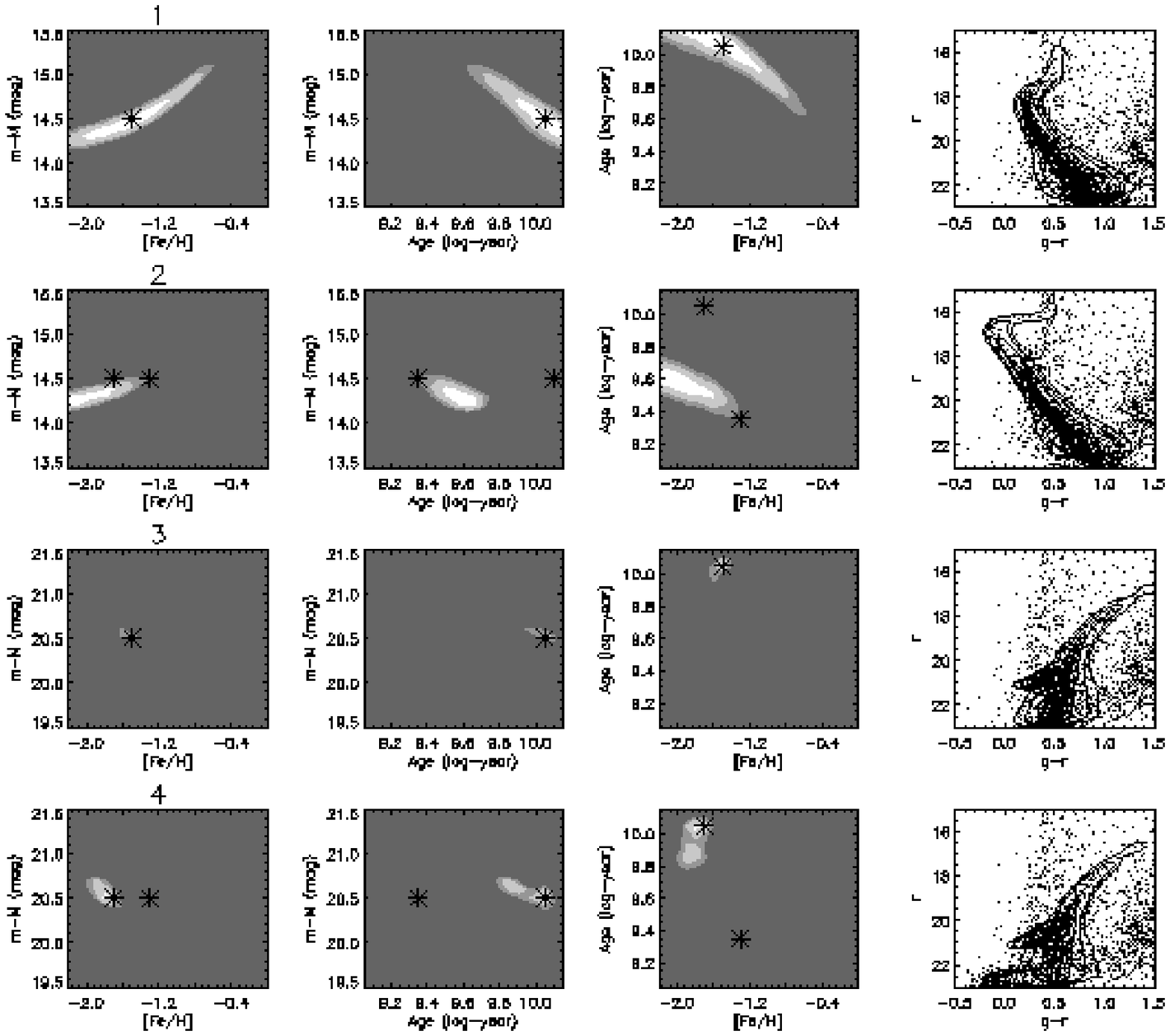}
\caption{Contour plots of the single component fitting results for the
  artificial CMDs. From left to right, the columns show: distance
  vs. metallicity; distance vs. age; age vs. metallicity; the
  CMD with the best-fit single component model in
  contours. The grey-scale contours show the regions in parameter
  space that are within 1$\sigma$, within 2$\sigma$, within 3$\sigma$
  and outside 3$\sigma$ of the best-fitting single component model,
  going from white to dark gray. Input values for the artificial
  stellar populations (see also Table \ref{tab:fakegcprops}) are
  indicated by asterisks.
}
\label{fig:fakescfit}
\end{figure*}

For the SC fits models are created for a grid of parameters
with a distance modulus range of 2.0 with 0.1 magnitude intervals
centered on the the actual distance modulus (Table
\ref{tab:fakegcprops}). Ages used run from 9.0 to 10.2 in log $t$ with
bin widths of 0.1 dex and metallicities go from [Fe/H]$=-$2.1 to 0.0
in steps of 0.1 dex and bin widths of 0.2 dex.
Each model plus the  ``control field'' is fit to an artificial CMD,
after which the resulting goodness-of-fit values are compared. The
resulting contour plots showing the best-fitting parameter
combinations are shown in Figure \ref{fig:fakescfit}.

\begin{figure*}[ht]
\epsscale{1.0}
\plotone{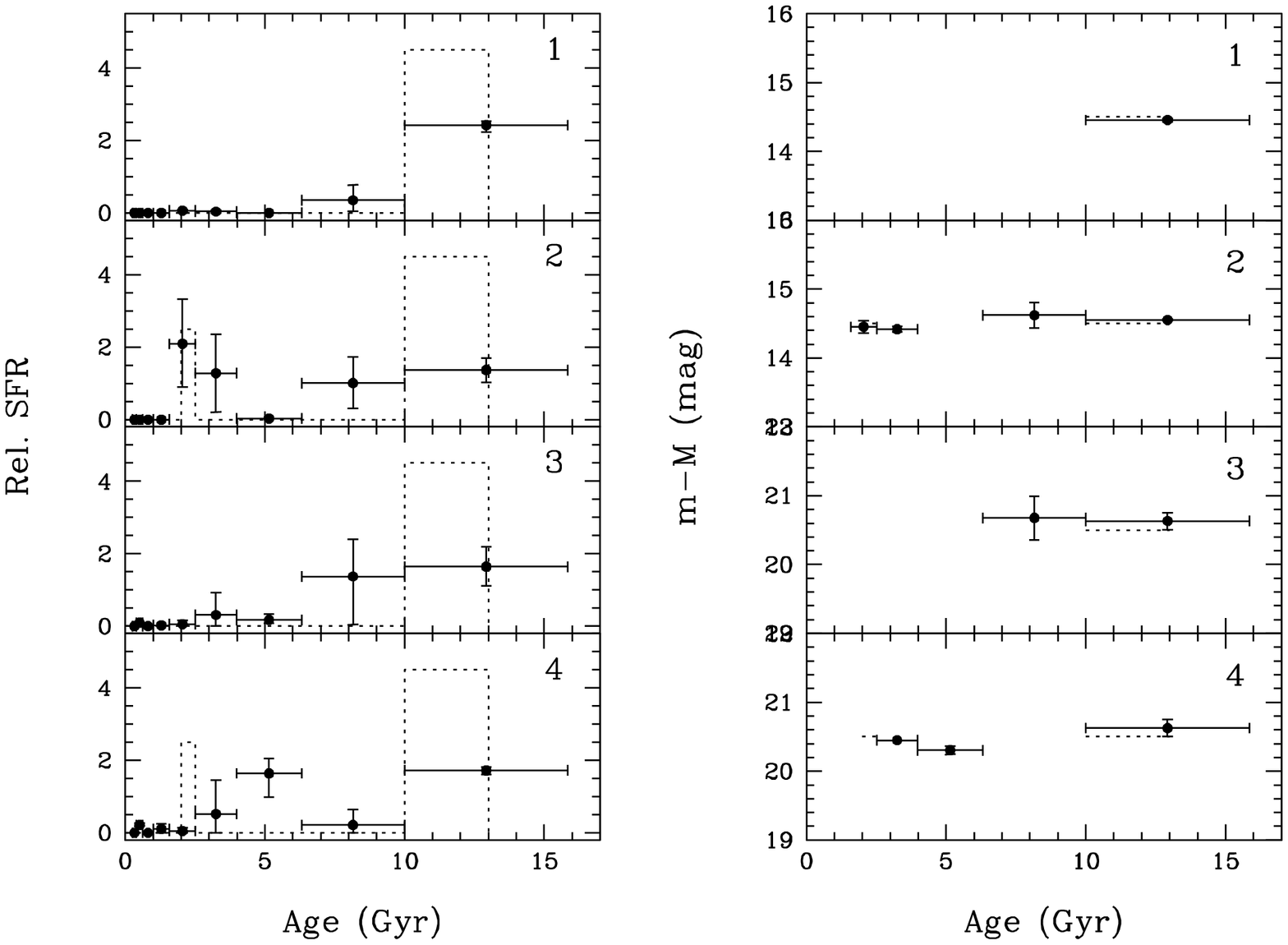}
\caption{Star formation histories and distances for the artificial
  CMDs from the ``distance solution'' mode. The columns on the left
  and the error bars are as in Figure \ref{fig:fakesfhamr}. In the
  columns on the right now the SFR-weighted average distance is
  plotted for the age bins with significant star formation. Dashed
  lines indicate the actual input distance for the artificial CMDs.  }
\label{fig:fakesfhdist}
\end{figure*}

Using the new distance solution option of MATCH we solve for the SFH
of the artificial CMDs again while fixing the metallicity to a single,
uniform value. The same age bins are used as for the SFH fits
described above, and a distance range of 3.0 is probed with 0.1
magnitude resolution. Including the ``control field'', 271 partial
CMDs are used in these fits. The fits are done at three different
values of the metallicity: [Fe/H]$=-$1.6, $-$1.5, and $-$1.4 to probe
the uncertainties. Figure \ref{fig:fakesfhdist} shows the resulting
SFHs in relative SFRs and distance determinations.

The SFHs in Figures \ref{fig:fakesfhamr} and \ref{fig:fakesfhdist}
show that the software succesfully recovers the single burst in
CMDs 1 and 2, and the two distinct episodes of star formation in CMDs
3 and 4. Some bleeding of star formation in adjacent bins will occur
even when the fitted age bins exactly match the input SFH
\citep{match}. Therefore it is to be expected that this also happens
here, since the duration of the input star forming episodes do not
match any age bins perfectly. It should be noted that the error bars
in the SFHs represent the range of SFRs encountered in
each age bin within the three fits on which the SFHs are based, which
means that the errors between adjacent bins are correllated. For
example, in the SFH of CMD 4 in Figure \ref{fig:fakesfhamr} the two
bins at ages of 2 to 4 Gyr are not independent; when the SFR in one of
them is low, it is high in the other. That the age found for the young
population in CMD 4 in Figure \ref{fig:fakesfhdist} is the natural
result of the metallicities in the fit being lower than the actual
metallicity of the younger population.
In all cases, the average metallicities and distances found in Figures
\ref{fig:fakesfhamr} and \ref{fig:fakesfhdist} agree well with the
input values.

Turning to the SC fits in Figure \ref{fig:fakescfit}, we see that for
CMDs 1 and 3, the input properties (indicated by stars) are recovered
very well. Because there is a degeneracy between distance, age and
metallicity when only the MS and MSTO are seen, the properties of CMD
1 are, however, constrained more poorly than for CMD 3. Since the SC
fits use a single model population, the results for CMDs with multiple
populations, such as CMDs 2 and 4, cannot be expected to properly
recover the properties of all stars. Rather, the population
representing the majority of stars in the CMD will have the strongest
weight. This is clearly visible in the SC fitting results for CMD 4,
that recover the properties of the older population, that dominates
the star counts in the RGB and HB. For CMD 2, however, the upper MS
and MSTO of the younger population has such high S/N that this
population does influence the SC fits significantly and the favored
age is relatively young ($\sim$4 Gyr).
This shows that for high S/N CMDs with a complex stellar population
make-up, the SC fits is not well suited; they were, however, developed
for application to low S/N CMDs.

What distinguishes these artificial CMDs from the real data, is that
the artificial stellar populations are based on the exact
same isochrones, IMF, and photometric error and completeness
model as used by MATCH. Inaccuracies in any of these will degrade the
fit results one obtains for real data compared to the results for the
artificial data.

\section{Tests on globular clusters}
\label{sec:gctests}

\begin{figure*}[ht]
\epsscale{0.8}
\plotone{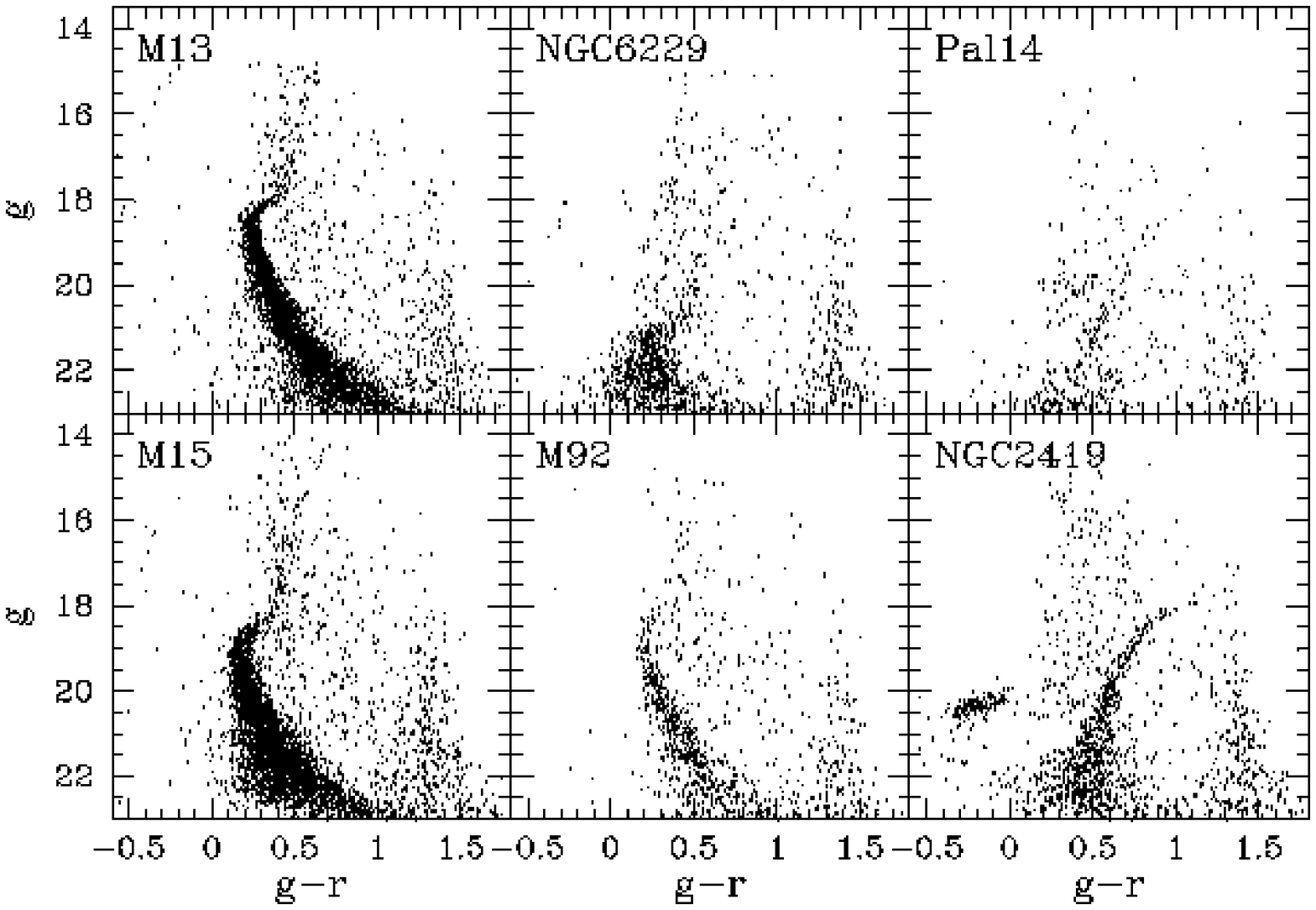}
\caption{Color-magnitude diagrams (CMDs) from SDSS DR5 data for the six
  globular clusters with independently determined population
  parameters, used to test the CMD-fitting software.
}
\label{fig:gccmds}
\end{figure*}

\begin{deluxetable*}{lccccc}
\tablecaption{Globular cluster properties}
\tablewidth{0pt} 
\tablehead{ \colhead{Cluster} & \colhead{R$_{\sun}$ (kpc)} &
  \colhead{m-M (mag)} &
  \colhead{[Fe/H] (dex)} & \colhead{Age (Gyr)} & \colhead{$<E(B-V)>$}}
\startdata
M 13 & 7.7, 7.5$^{a)}$ & 14.38, 14.43 & $-$1.54 & 12$^{a)}$ & 0.019 \\
M 15 & 10.3, 10.0$^{b)}$ & 15.37, 15.0$^{b)}$ & $-$2.26 & 13.2$^{b)}$ & 0.11 \\
M 92 & 8.2 & 14.6 & $-$2.28 & 14$^{c,d)}$ & 0.025 \\
NGC 2419 & 84.2, 98.6$^{e)}$ & 19.63, 19.97$^{e)}$ & $-$2.12 & 14$^{e)}$ & 0.060 \\
NGC 6229 & 30.4 & 17.44 & $-$1.43, $-$1.1$^{f)}$ & $\sim$11.5$^{g)}$ & 0.023 \\
Pal 14 & 73.9 & 19.34 & $-$1.52 & $\sim$10$^{h)}$ & 0.035 \\
\tablecomments{Values are from \cite{harris96}, except for: a)
  \cite{grundahl98}; b) \cite{mcnamara04}; c) \cite{pont98}; d)
  \cite{grundahl00} ; e) \cite{harris97}; f) \cite{borissova99}; 
  g) \cite{vandenberg00}; h) \cite{sarajedini97} }
\enddata
\label{tab:gcproperties}
\end{deluxetable*}

Globular clusters are ideal test cases for the SDSS implementation of
MATCH since in most of them their stars have been found to be of a
single age and nearly single metallicity. Here we will use six GCs
present in the SDSS data to test how well MATCH, using the isochrones
from \cite{girardi04}, can recover their known population
properties. 
The systems in question, \object{M13}, \object{M15},
\object{M92}, \object{NGC 2419}, \object{NGC 6229}, and \object{Pal
14}, cover a large range in distance, background contrast (i.e. ratio
of member stars versus field stars) and have metallicities between
[Fe/H]$=-2.3$ and $-1.3$ dex.  For the following tests, all stars
within a 5\degree$\times$5\degree~ field centered on each GC were
obtained from the SDSS catalog archive server (CAS). It should be
noted that in the cases of M15, M92 and NGC 2419 these fields were not
completely covered by the DR5 footprint. Flags were used to exclude
objects with poor photometry\footnote{See
http://cas.sdss.org/astro/en/help/docs/realquery.asp\#flags} and all
stars were individually corrected for foreground extinction using the
correction factors provided by DR5 and based on the maps by
\cite{sfd}. Any contamination of the samples of stars obtained by
misclassified galaxies or other contaminants do not pose problems for
our analysis because of the use of control fields that should contain
the same populations of these contaminants.  ``Control field'' CMDs
were constructed from all stars outside a radius of 0.5\degree~ from
the center of each GC. ``Target fields'' were constructed by selecting
stars within a certain radius from the center.  The radii used for
M13, M15, M92, NGC 2419, NGC 6229, and Pal 14 are 0.25\degree,
0.2\degree, 0.25\degree, 0.15\degree, 0.15\degree, and 0.1\degree,
respectively.  These radii are similar to the tidal radii of the GCs
but are not strictly related to them, since in all clusters -- and
especially in the densest (M13 and M15) -- the SDSS photometry in the
central parts is missing due to the high degree of crowding. With the
radii used, a significant number of field stars is present in each
target CMD, so that MATCH has a good handle on the proper scaling of
the control CMD.

The SDSS CMDs of the GCs are shown in Figure \ref{fig:gccmds} and
their previously measured parameters (distance, age, [Fe/H]) are
listed in Table \ref{tab:gcproperties}; the mean reddening $E(B-V)$ in
magnitudes according to the \cite{sfd} maps in the target area for
each GC is also listed in the table. The GCs in the three top panels
in Figure \ref{fig:gccmds} have metallicities of around [Fe/H]$=-1.5$
dex, and the GCs in the lower panels are more metal-poor
([Fe/H]$<-2.0$). For the more nearby systems (M13, M15, M92 and NGC
6229) the main CMD features that will drive the fits are the main
sequence (MS) and main sequence turn off (MSTO); for the more distant
systems (NGC 2419 and Pal 14) only the red giant branch (RGB) and
horizontal branch (HB) are observed.  Pal 14 in particular represents
the case of a very sparsely populated system with high contamination
from field stars.

In the following subsections, we discuss the results of applying MATCH
to these GCs in the four different modes described in section
\ref{sec:modes}. In fitting the CMDs of these objects we will use a
color range of $-0.5<g-r<1.5$, and limit the magnitude range to
$g<22.5$ and $r<22.0$; the color cut includes the MSTO and the tip of
the RGB, but excludes a large fraction of the numerous low-mass
foreground stars.  For all the fits in this section we will assume a
slope for the stellar initial mass function that is close to Salpeter;
the binary fraction is assumed to be 0.1, in line with recent
observational and theoretical evidence for low binary fractions in
globulars \citep[e.g.,][]{ivanova05,zhao05}.

\subsection{Star formation histories}
\label{subsec:gcsfh}

\begin{figure*}[ht]
\epsscale{1.0}
\plotone{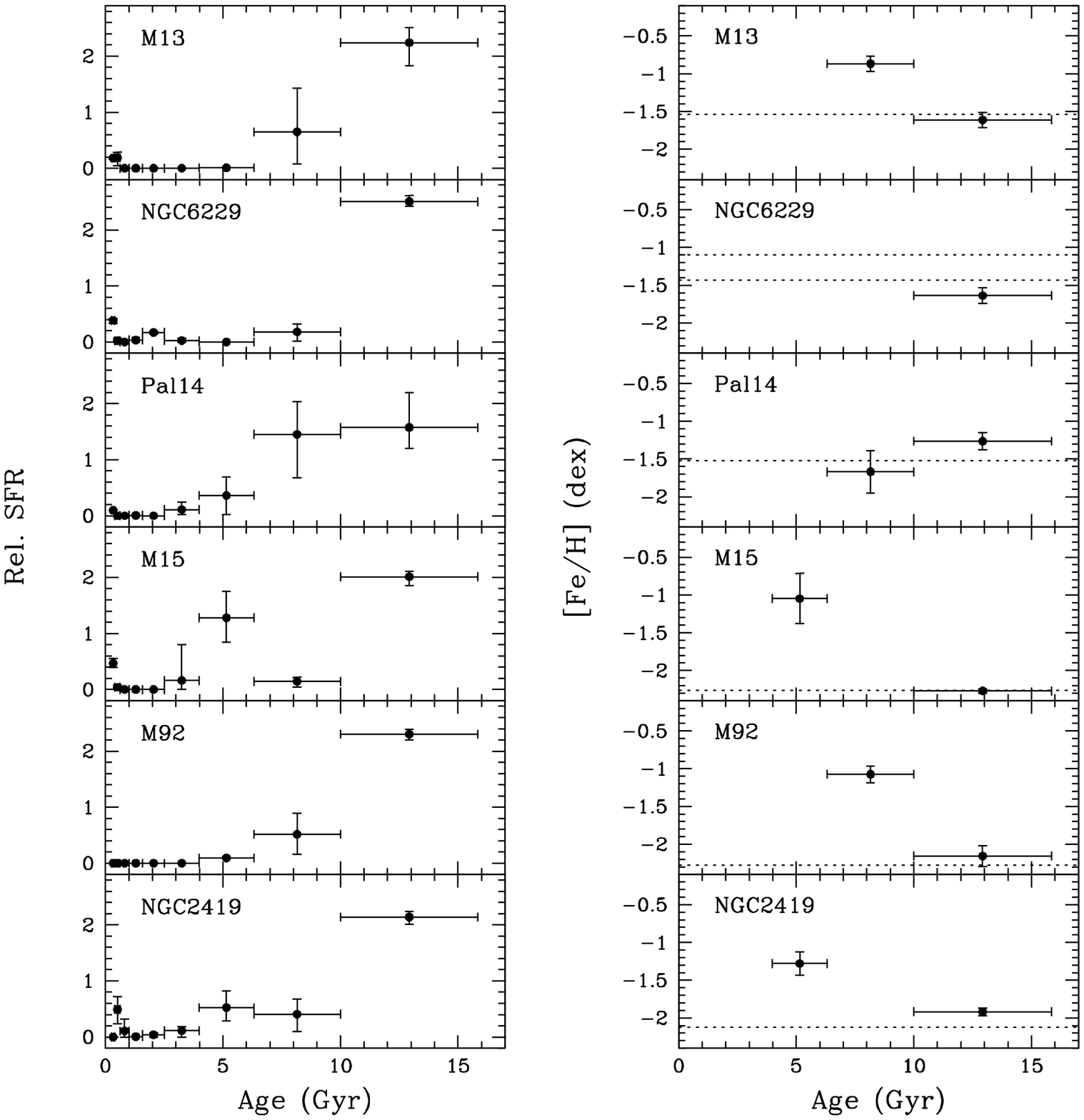}
\caption{Star formation histories and age metallicity relations for
  the six globulars obtained by using MATCH in its ``standard'' mode
  (fitting for age and metallicity distributions at a given
  distance). Star formation rates are shown relative to the average
  star formation rate over the entire age range. Error bars along the
  age axis show the width of each age bin, while the error bars along
  the y-axis show the range of star formations in the age bin at the
  three different distances for which a fit was done and the
  uncertainty in the metallicity in each age bin. Metallicities are
  only shown for bins with significant star formation. The dashed
  lines in the right panels indicate the literature values for the
  metallicity from Table \ref{tab:gcproperties}. }
\label{fig:gcsfh}
\end{figure*}

\begin{figure*}[ht]
\epsscale{1.0}
\plotone{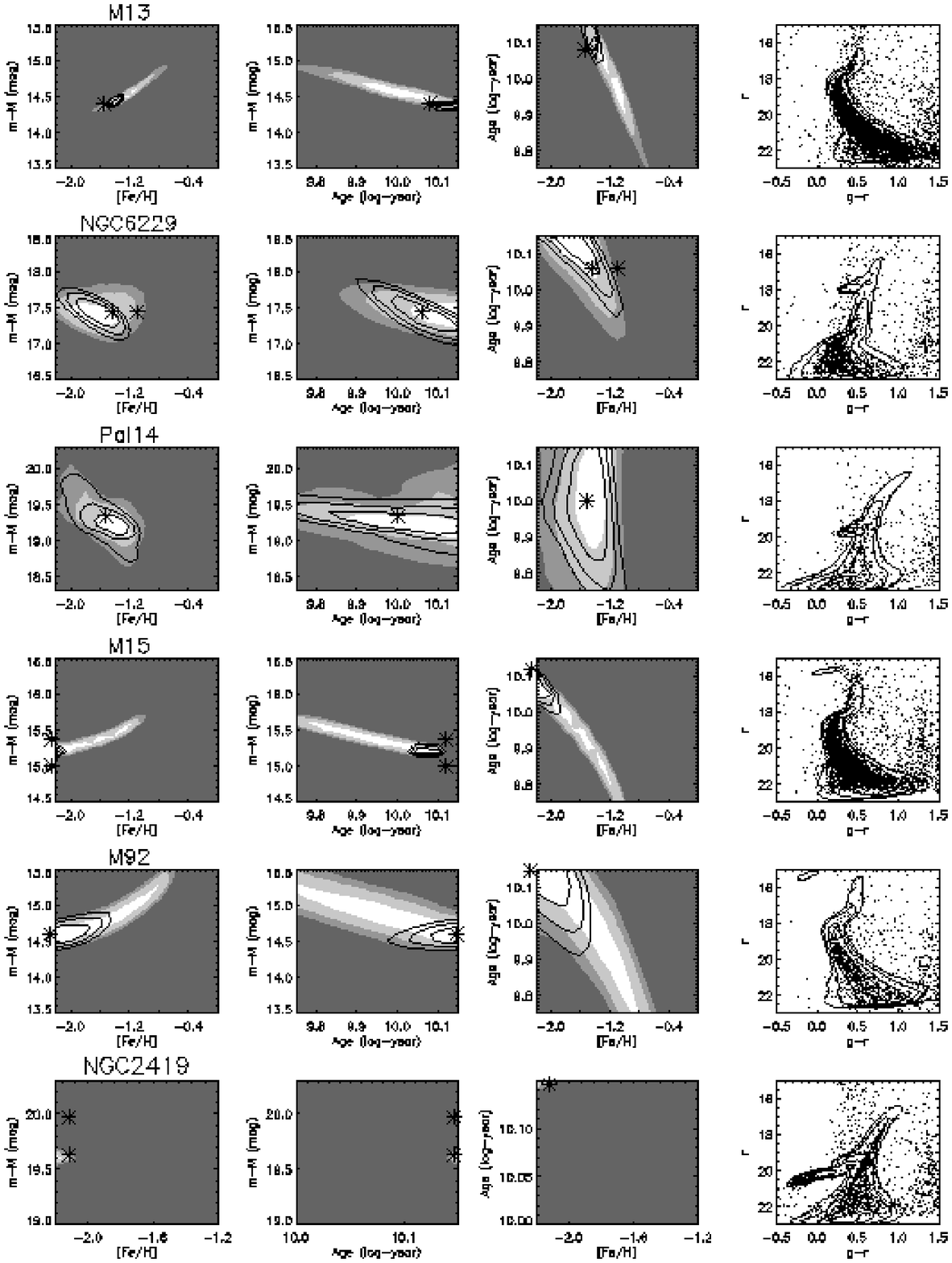}
\caption{Comparison of single component population parameters from
  SDSS and MATCH versus their literature values for the six
  globulars. The columns show (from left to right): distance
  vs. metallicity; distance vs. age; age vs. metallicity; the
  color-magnitude diagram with the best-fit single component model in
  contours. The grey-scale contours show the regions in parameter
  space that are within 1$\sigma$, within 2$\sigma$, within 3$\sigma$
  and outside 3$\sigma$ of the best-fitting single component model,
  going from white to dark gray. The solid line contours are for
  the case in which one of the parameters is fixed to its best-fit value,
  i.e. age for the left-most, metallicity for the middle-left, and
  distance for the middle-right columns. Literature values for the
  parameters are indicated by asterisks.  }
\label{fig:gcscfit}
\end{figure*}

To measure the SFH and AMR of the clusters we run MATCH fits with nine
age bins and twelve metallicity bins. Because the isochrones are spaced
more evenly in log $t$ than in $t$, the age bins are defined to have
equal width in log $t$, with the oldest running from 10.0 to 10.2 (10
Gyr to 15.8 Gyr) and the youngest from 8.6 to 8.8 (400 Myr to 630
Myr). The metallicity bins cover the range from [Fe/H]$=-$2.4 to 0.0
in twelve bins of 0.2 dex width. Together with the control CMD
this yields 109 model CMDs for each GC. The fitting
software then finds the linear combination of these model CMDs that
best fits the observed CMD.  Each object is fit several times with
slightly different fixed distances to determine the impact of distance
errors on the recovered solutions. In the cases of M15 and NGC 2419, where
values from the literature for the distance modulus are quite discrepant,
we run fits for distance moduli of 15.0 to 15.4 (M15) and 19.6 to 20.0
(NGC 2419) in 0.1 magnitude steps. For the other clusters we run three
fits each, one at the literature value and at 0.1 magnitude further
and nearer.

Figure \ref{fig:gcsfh} shows the results of these fits, with the
fractional SFR as a function of time in the left
column, and the metallicity as a function of time in the right
column. Metallicities are only plotted for age bins with a significant
SFR.  The plotted SFR for a given age bin is the average of the
values given by the fits at the different distances, with the error
bars showing the complete range of SFR values found. The metallicity
in a bin is the average of the fits  weighted by the SFR, and the
error bars correspond to the standard deviation.

Looking at the three intermediate metallicity GCs in the top half of
Figure \ref{fig:gcsfh} we see that in all cases MATCH infers a purely
old ($>$7 Gyr) population, as expected.  For M13 and NGC 6229
practically all stars are found to be older than 10 Gyrs; some
bleeding into the second oldest age bin is always likely to occur,
cf. Section \ref{subsec:fakegc} and \cite{match}.  In the case of Pal
14 the SFR is found to be constant over the two oldest age bins, but
this only confirms the age estimate of $\sim$10 Gyr
\citep{sarajedini97}, which implies that star formation would likely
have been spread over these two bins.
The intermediate metallicities of M13, NGC 6229 and
Pal 14 are reproduced well and our recovered values are within 0.1 to
0.2 dex of the literature values. For NGC 6229 our fits favor the
slightly lower metallicity from \cite{harris96} rather than the higher
value from \cite{borissova99}. 

Turning to the bottom half of Figure \ref{fig:gcsfh} with the three
metal-poor clusters, we see that here also the results are consistent
with a purely old population, although the SFHs of M15 and NGC 2419
show a non-zero SFR in a few younger age bins. In the case of M15 this
is due to the blue horizontal branch (BHB) stars that are not properly
reproduced by the theoretical isochrones, causing the code to fit them
as a younger turn-off. This demonstrates that CMD fitting results
always need to be looked at in detail in order to understand their
implications.
The age distribution in NGC 2419 is not very
well constrained because no MSTO is present in the CMD.
The metallicities of M15 and M92 are recovered well for the old stars.
While the estimated metallicity of $-$1.9 for NGC 2419 is on the high
side, for these metal-poor GCs the recovered metallicities are within
0.2 dex of the literature values.

\begin{figure*}[ht]
\epsscale{1.0}
\plotone{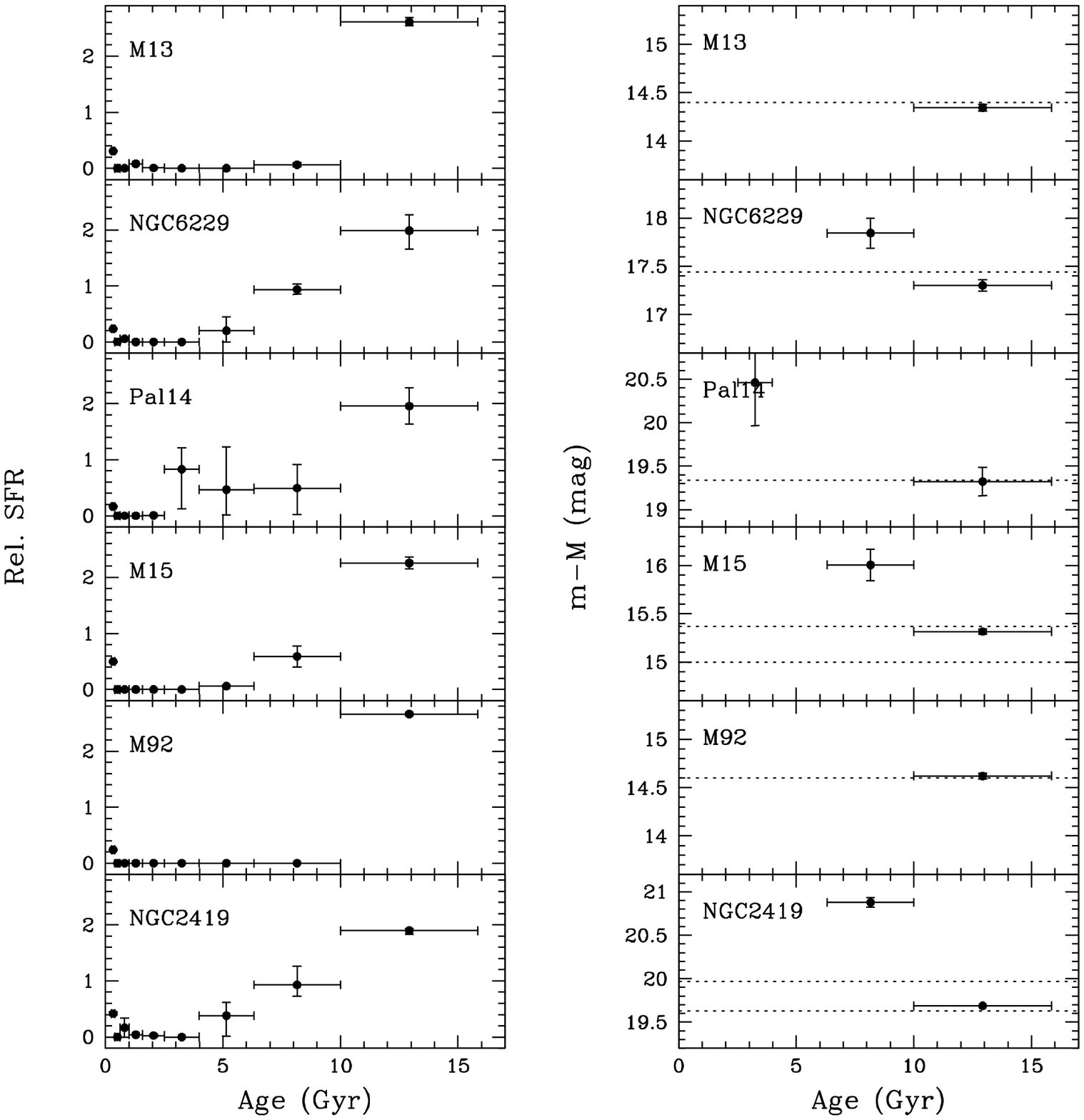}
\caption{Star formation histories and distances for the six globulars
  obtained by using MATCH in the distance solution mode (Section
  \ref{subsec:distsol}). As in Figure
  \ref{fig:gcsfh} star formation rates are shown relative to the
  average star formation rate over the entire age range and horizontal
  error bars show the width of the age bins. Vertical error bars
  indicate the range of star formation rates encountered at the three
  different metallicities for which a fit was done, and the
  uncertainty of the distance. The dashed
  lines in the right panels indicate the distance values from the
  literature (see Table \ref{tab:gcproperties}).
}
\label{fig:gcdfit}
\end{figure*}

\subsection{Single component fitting}
\label{subsec:gcscfit}

\begin{figure*}[ht]
\epsscale{1.0}
\plotone{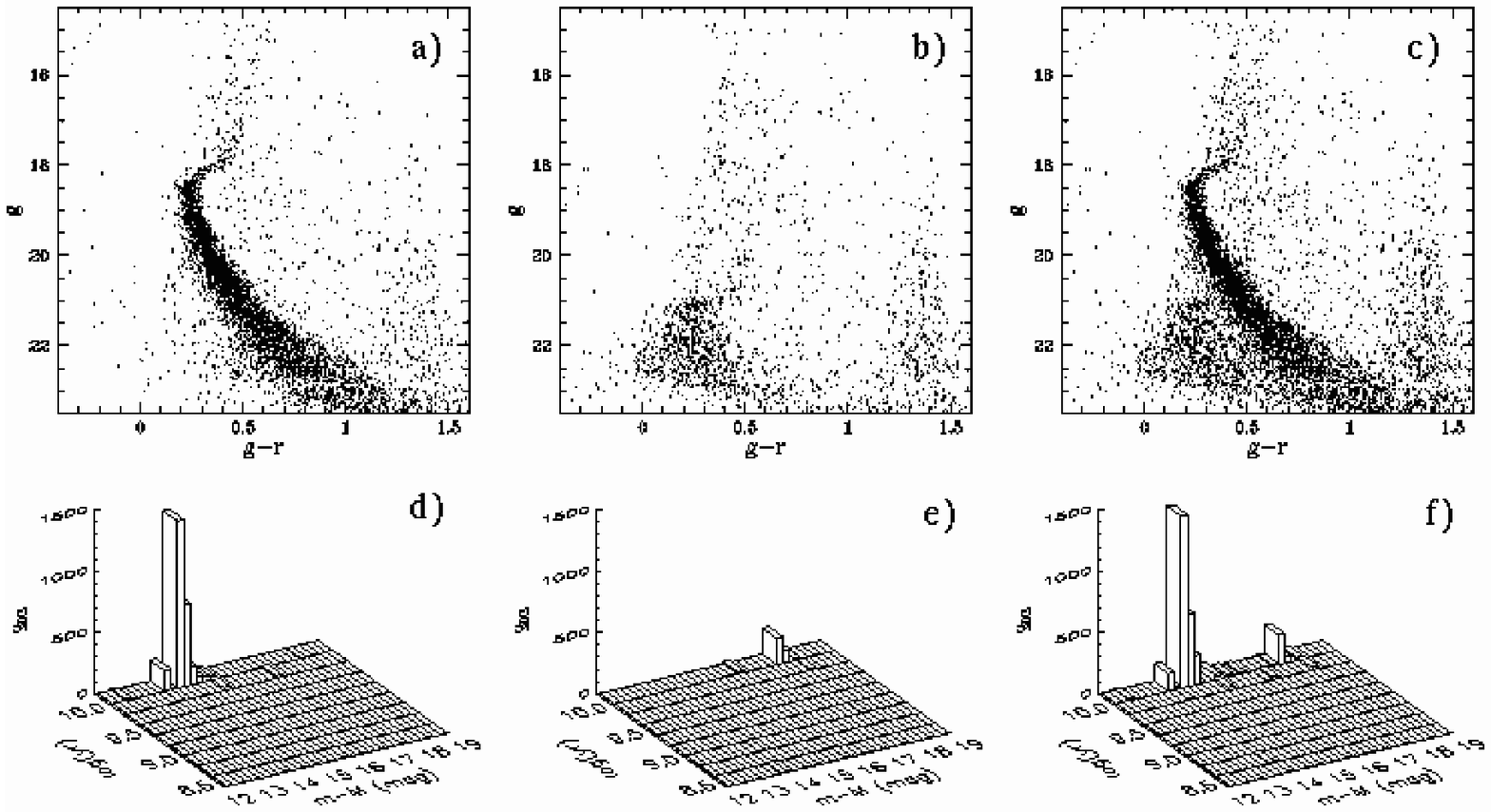}
\caption{MATCH analysis of a composite CMD: {\it a)} CMD of the
  globular M 13. {\it b)} CMD for the globular NGC 6229. {\it c)}
  combined CMD of the globulars M 13 and NGC 6229.  {\it d)} average
  result of the three distance solution fits to the CMD in panel a),
  showing the number of stars in each age-distance bin; the M 13 stars
  are fit by the oldest age bins around m-M=14.4 magnitudes. {\it e)}
  as in d) but for the CMD in panel b); the NGC 6229 stars are fit by
  the oldest age bins at m-M=17.2 and 17.4 magnitudes. {\it f)} as in
  d) and e) but for the fits to the combined CMD in panel c); both the
  ages and distances of the two distinct populations are correctly
  recovered.  }
\label{fig:m13ngc6229}
\end{figure*}

As described in section \ref{subsec:scfit} above, in the Single Component
(SC) fitting mode an individual model CMD is fit to the data together
with the control CMD, rather than a whole set of models. The results
therefore provide information on the characteristic properties of the
dominant stellar population in the CMD.  We sample the 3D parameter
space of distance, age and metallicity by using one of nine age bins,
one of 21 distance bins, and one of 23 metallicity bins. The age bins
have a width of 0.1 log-years, going from 10.1-10.2 log-years (12.5
Gyr to 15.8 Gyr) to 9.7-9.8 log-years (5 Gyr to 6.3 Gyr) in steps of
0.05 log-years. This age range more than suffices for these GCs. We
use a distance modulus range of two magnitudes centered on the
literature value with a resolution of 0.1 magnitude. Metallicity bins
are centered at [Fe/H]$=-$2.2 to 0.0 with an inter-bin distance of 0.1 dex
and a width of 0.2 dex.

After matching each of the 4347 SC options to each GC and determining
the expected variance in the goodness-of-fit parameter $Q$, we take
the best fit as our measurement and compare the other fits with
respect to these best values. The resulting likelihood contour plots
for metallicity, distance and age measurements are shown in Figure
\ref{fig:gcscfit}. Note that in these contour plots, the third
parameter is unconstrained, rather than fixed at its known value. If
the third parameter would be kept fixed, the contours would shrink
significantly.  In general the agreement with the literature values
(indicated with asterisks in Figure \ref{fig:gcscfit}) is very good.
The small offset for M13 is most likely due to inaccuracies in the
isochrones, which illustrates the added uncertainty in CMD fitting
results that depend on a single set of theoretical isochrones. Our
distance modulus measurement from the SC fits for M15 lies exactly in
between the two literature values, while for NGC 2419 our results
clearly favor the smaller distance value from \cite{harris97}. From
the two metallicity determinations in the literature for NGC 6229, the
fit results agree significantly better with the lower metallicity of
[Fe/H]$\sim -$1.4 dex.  In some cases, it looks as if the peak of the
likelihood lies outside of the parameter space probed, at lower
metallicities or older ages. Unfortunately, these limits of the
parameter space are dictated by the parameter range of the available
isochrones.

Also shown in the
right column of Figure \ref{fig:gcscfit} are the observed CMDs of the
GCs, with contours representing the model populations. In most cases
we used the literature parameters for these models, except for M15,
where we chose a distance modulus of 15.2, the mean of the
two different available values. For NGC 6229 and
NGC 2419 we use the literature metallicities and distance moduli that best
correspond to our results. The agreement of model and data is
generally acceptable, although inaccuracies, especially in the
sub-giant branch, can be seen in M13 and M15.

Important to note is the influence of the degeneracies between age,
distance, and metallicity. For the nearby systems, M13, M15 and M92,
where the fit is driven by the MS, there is a clear
distance-metallicity covariance; the left-most panels in Figure
\ref{fig:gcscfit} show that the distance estimate increases with
increasing metallicity. For the distant GC Pal 14, where the fit is
driven by the RGB, the covariance is the other way around.  Another
important, yet not wholly surprising, point is that higher S/N in the
CMD features results in much tighter contours, illustrated by the fact
that M13, M15 and NGC 2419 give the best constrained results. NGC 2419
has the additional advantage that the well-populated HB provides a
very strong constraint on the distance, making this system the
best-constrained of the lot.

\subsection{Distance solutions}

As mentioned in Section \ref{subsec:distsol}, we explore fits for
different distances and different ages, with the metallicity linked to
the age.  The result of each fit is a SFH with a (potentially
different) distance for each age bin; differing distances for
different sub-populations are not astrophysically relevant for
globular clusters, but serve as a test case.  For fitting the GCs it
seems reasonable to use the same metallicity in all age bins.  We use
the same nine age bins as in Section \ref{subsec:gcsfh}. In distance
space we use a distance modulus range of three magnitudes centered on
the known distance of each system with a resolution of 0.1 magnitudes,
giving a total of 31 distance bins.  For each GC we carry out three
fits with different metallicities, centered on the literature values
(see Table \ref{tab:gcproperties}) with offsets of 0.1 dex. For
NGC 6229 we assume the lower metallicity value [Fe/H]$=-$1.43, which is
favored by our SFH and single component fitting results.

In Figure \ref{fig:gcdfit} we present the age-distance distributions
that result from these fits. Compared to the fits in section
\ref{subsec:gcsfh} the number of free parameters per fit is the same,
with now metallicity fixed instead of distance. Thus, in general
similar quality SFHs should be expected, although not necessarily
exactly the same. While in the ``standard'' SFH fits the star
formation within an age bin can be spread over several metallicity
bins, it can now be spread over several distance bins. These spread
will translate differently into the color-magnitude plane, and
therefore can result in differences between the SFHs in Figures
\ref{fig:gcsfh} and \ref{fig:gcdfit}.

As in Figure \ref{fig:gcsfh} all GCs are found to have a single peak
in their SFR in the oldest age bin, in some cases with some bleeding
into the adjacent bin. The extension of the SFH of Pal 14 to more
recent times has large uncertainties and is not significant. It should
be taken into account that the exact age of this system is difficult
to constrain with SDSS data, since the MSTO is not present.
Interesting to see is that M13 and M92 show almost no bleeding into
the second oldest age bin here, while they did in Figure
\ref{fig:gcsfh}; for M15 and NGC 6229 the exact opposite is the
case. Apparently, the former are fit better with some distance spread,
while the latter benefit more from a metallicity spread. This does not
necessarily reflect an actual distance or metallicity spread, but
rather shows that inaccuracies in the isochrones are sometimes better
accounted for one way and sometimes another way.

The distances we find for the oldest age bin, and
therefore for the majority of the stars, show very good agreement with
the literature values. Comparing Figures \ref{fig:gcsfh} and
\ref{fig:gcdfit} shows that the SFHs obtained when fixing either the
distance or the metallicity are very similar.

The main motivation for enabling this distance fitting mode was,
however, not to analyze single-distance systems, but to
be able to cope with cases where stars are located at different
distances \citep[cf. Canis Major, ][]{dejong07}. To demonstrate this
application, we combine the CMDs of M13 and NGC 6229, resulting in a
CMD containing two populations with similar metallicities and ages,
but located at significantly different distances. The CMDs of M13, NGC
6229 and the combined CMD are shown in panels a), b) and c) of Figure
\ref{fig:m13ngc6229}. We run distance solution fits to these CMDs in
the same way as before, with the same age bins, but now using a larger
distance modulus range, from 12.0 to 19.0 magnitudes with a resolution
of 0.2 magnitudes. Three fits are run, for uniform metallicities of
[Fe/H]$=-$1.4, $-$1.5, and $-$1.6.  We include also a control CMD in
the fit which is the control CMD of M 13; since M 13 and NGC 6229 are
only 10 degrees apart and at similar Galactic latitudes, this gives
good results. The averaged best-fitting distributions of stars over the grid
of age and distance bins are shown in panels d), e) and f) of Figure
\ref{fig:m13ngc6229}. Both ages and distances of the two populations
are recovered extremely well, with all stars in the oldest age bin ($>$10
Gyr), M13 stars centered at a distance modulus of 14.4
and NGC 6229 stars at a distance modulus of 17.4. Moreover, the
result for the combined CMD is almost exactly the sum of the results
for the individual CMDs, clearly illustrating the power of MATCH to
decompose more complex CMDs.

\subsection{Over-density detection}

\begin{figure*}[ht]
\epsscale{1.0}
\plotone{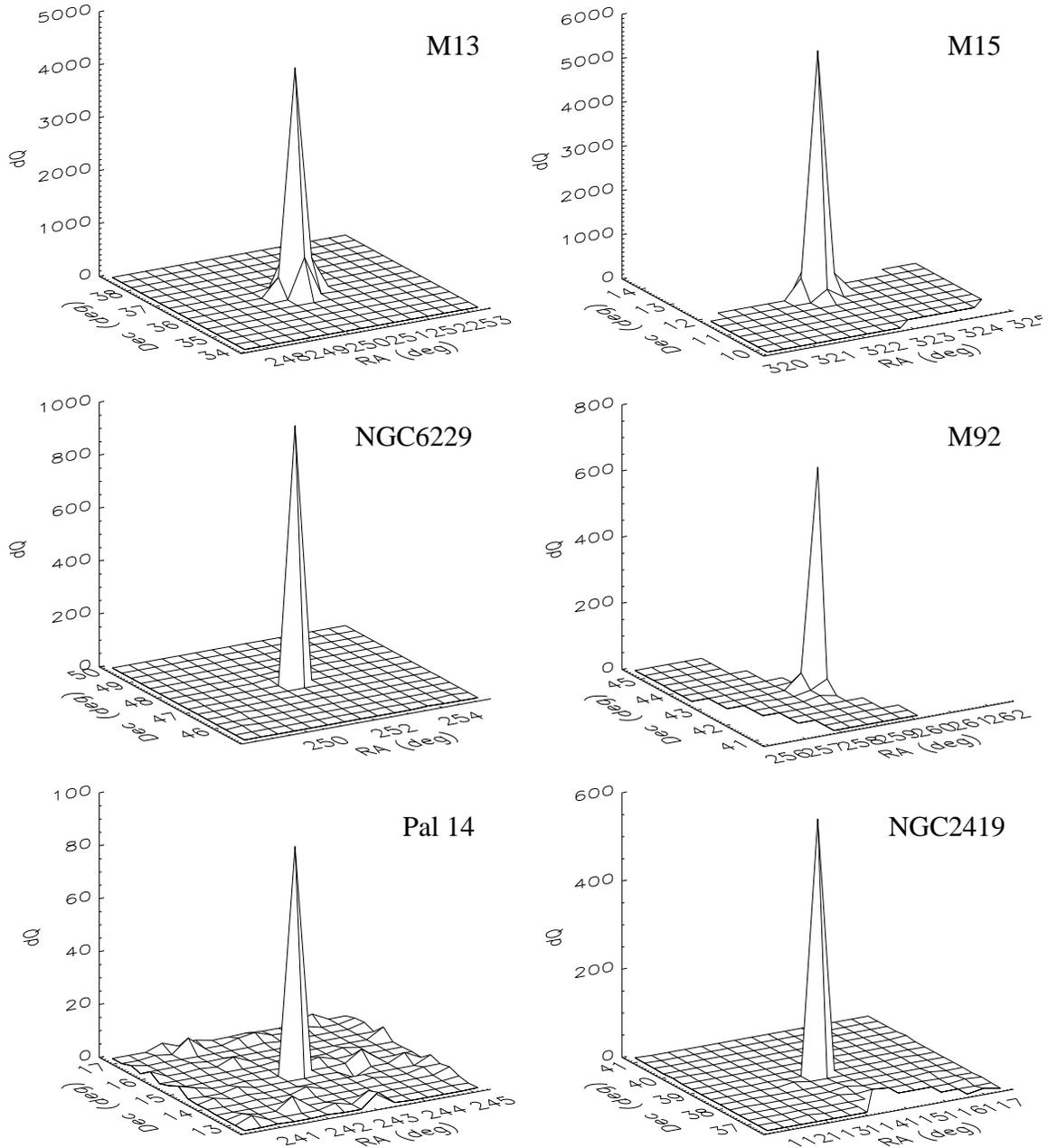}
\caption{Stellar overdensity detection using MATCH: each panel shows
  the fit quality improvement, d$Q$, between fitting only a control
  field and fitting a control field plus stellar population model to
  the CMDs of 20\arcsec$\times$20\arcsec regions around the six
  globular clusters. The fields plotted are 5\degree$\times$5\degree
  fields centered on the globulars with right ascension along the
  x-axis.  }
\label{fig:gcareas}
\end{figure*}

The final application we describe and test here is aimed at
algorithmically detecting stellar populations which are confined to a
limited distance and area on the sky, i.e. that form
``overdensities''. Of course, the GCs used in this section are very
obvious stellar overdensities and should be detected at very high
S/N. Following the approach described in section \ref{subsec:detect}
we define a grid of 20\arcmin$\times$20\arcmin~ boxes in each
5\degree$\times$5\degree~ field that was obtained from the SDSS
catalog server, with the central box centered on the GC. The CMD of
each box is fit with the control field CMD and with the control field
plus a simple population model.  For each GC we use a model with the
appropriate distance, age and metallicity. Figure \ref{fig:gcareas}
shows the difference in the goodness-of-fit parameter $Q$ between the
two fits for areas around each GC. Note that for M15 and M92, the
5\degree$\times$5\degree~ field is not completely observed by SDSS.
Clearly, all GCs produce high peaks in $\Delta Q$. Since $\Delta Q$ is
roughly proportional to the number of stars attributed to the
population model in the model fit, the objects with the largest
numbers of stars produce the strongest detections. Only in the plot
for Pal 14 does the vertical scale allow the random variations in
$\Delta Q$ to be visible in the area around the object. The S/N of
these detections should be measured with respect to this random
noise. Using the 216 boxes around the central 9 boxes we obtain an
estimate of the rms in $\Delta Q$ from random variations in the
CMDs. Negative values of $\Delta Q$ are suppressed because a negative
number of stars in a population is unphysical, which has to be taken
into account when calculating the rms. Even Pal 14, a rather sparsely
populated and distant cluster, is detected with high S/N, as $\Delta
Q_{Pal14}/rms_{\Delta Q} \simeq 56$.

\newpage

\section{Application to new Milky Way satellites}
\label{sec:newsats}

In summary, Section \ref{sec:gctests} has illustrated the robustness
and precision with which MATCH can interpret high-quality CMDs in a
variety of circumstances.  Over the past few years the importance of
deep wide-field surveys for the study of the stellar components of the
Milky Way galaxy has been demonstrated by 2MASS
\citep[e.g.,][]{ojha01,ibata02,majewski03,brown04,martin04} and the
SDSS. Specifically, the SDSS has enabled a detailed picture of the
Galactic stellar halo \citep[e.g.,][]{fieldofstreams,bell07} and led
to the discovery of at least fourteen previously unknown dwarf
galaxies and globulars: \object{Ursa Major I} \citep[UMaI,][]{UMaI},
\object{Willman 1} \citep[Wil1,][]{Wil1}, \object{Canes Venatici I}
\citep[CVnI,][]{CVnI}, \object{Bootes I} \citep[BooI,][]{Boo},
\object{Ursa Major II} \citep[UMaII,][]{UMaII}, \object{Coma
Berenices}, \object{Canes Venatici II}, \object{Hercules}, \object{Leo
IV}, \object{Segue 1} \citep[Com, CVnII, Her, LeoIV, Seg1,][]{5pack},
\object{Leo T} \citep{LeoT}, \object{Bootes II}
\citep[BooII,][]{BooII}, and two very low-luminosity globulars
\object{Koposov 1} and \object{Koposov 2} \citep{koposov07a}.  While
these discoveries are important for our understanding of the so-called
``missing satellite problem'' \citep[e.g.,][]{klyp99,moore99}, these
new systems might also shed new light on the mechanisms of star
formation in low-mass halos with small baryon content. Why these
systems were only discovered with SDSS data can be explained by their
low surface brightness, as shown by \cite{koposov07b}.  Apart from the
promise that more systems of even lower surface brightness might be
waiting to be discovered by future generations of deep imaging
surveys, this fact raises intriguing questions.  Do these systems show
that star formation can occur under conditions generally considered to
be incapable of supporting such activity, or have these systems
evolved to their current state through, for example, tidal
interactions with the Galaxy? A thorough and systematic study of the
properties of these objects is needed to understand their formation
and evolution and the possible cosmological implications.

Several groups have been working on photometric and spectroscopic
follow-up of the new MW companions in order to constrain their
properties. A compilation of these results is given in Table
\ref{tab:newprops}. Although individual measurements of properties might
be less precise in some cases, the advantage of studying these objects
using the SDSS photometry is that this highly uniform dataset allows
a systematic approach to the complete set of objects and fair
comparisons between them. In the following, we apply our single
component (SC) fitting techniques to constrain the overall distances,
metallicities and ages of the objects listed in Table
\ref{tab:newprops}. We also constrain the SFH of a subset of objects,
since the majority contain too few stars to make this a worthwhile
exercise. The objects we use are Boo, CVn I, and UMa I because of
their relatively well-populated CMDs, and Leo T because of its clearly
complex star formation history. For all the fits in this section we
assume a Salpeter-like initial mass function and a binary fraction of
0.5, similar to the value for disk stars \citep{duquennoy91,kroupa93}.

\begin{figure*}[ht]
\epsscale{0.8}
\plotone{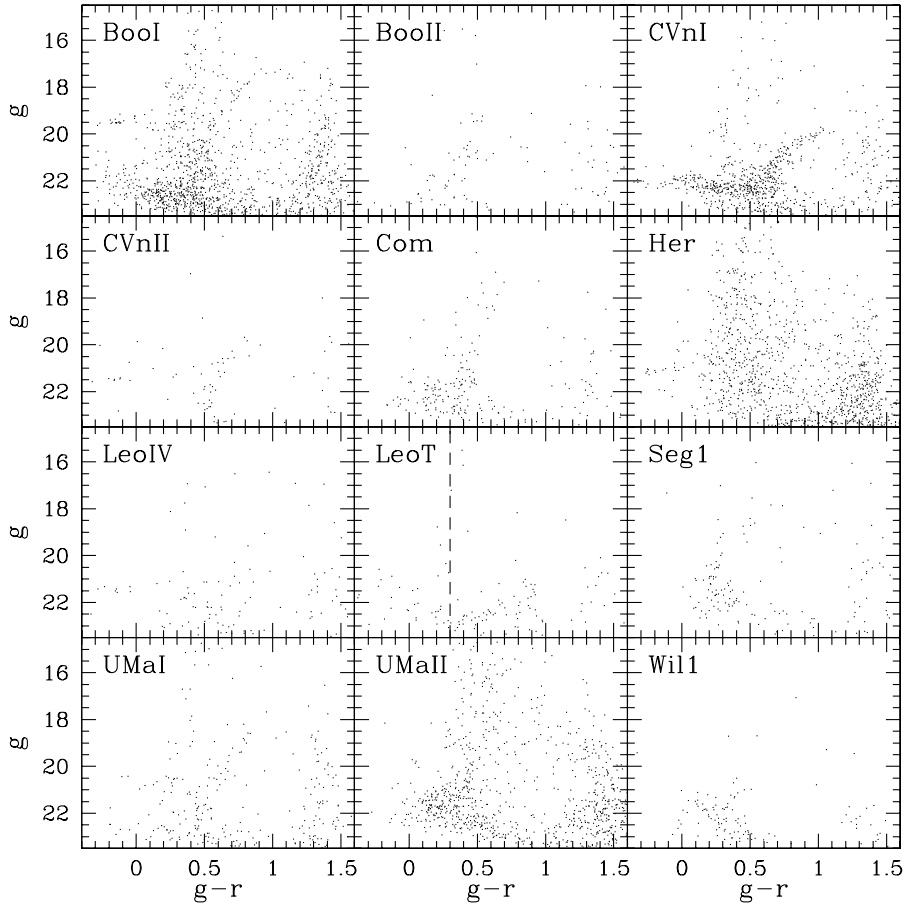}
\caption{Color-magnitude diagrams from SDSS data for twelve recently
  discovered Milky Way satellites. The dashed line in the panel for
  Leo T indicates the separation between the red and blue parts used
  in the single component fitting analysis (see Section
  \ref{subsec:nsscfits}).  }
\label{fig:nscmds}
\end{figure*}

\thispagestyle{empty}
\begin{deluxetable*}{lccccc}
\tablecaption{New MW satellite properties}
%\rotate
\tablewidth{0pt} 
\tablehead{ \colhead{Object} & \colhead{R$_{\sun}$ (kpc)} &
  \colhead{m-M (mag)} &
  \colhead{[Fe/H] (dex)} & \colhead{$r_h$} & \colhead{$M_V$ (mag)} }
\startdata
Bootes I & 60$\pm$6  & 18.9$\pm$0.2, 19.1$\pm$0.1$^{a)}$, 19.0$\pm$0.1$^{b)}$ & $\sim$-2, $\sim$-2.5$^{c)}$, $\sim$-2.1$^{d)}$  & 12\arcmin.8$\pm$0\arcmin.7  & -5.8$\pm$0.5  \\
Bootes II & 60$\pm$10 & 18.9$\pm$0.5 & $\sim$-2 & 4\arcmin$\pm$2\arcmin & -3$\pm$1 \\
Canes Venatici I & 220$\pm$20 & 21.75$\pm$0.20  &  $\sim$-2, -2.5 -- -1.5$^{e)}$, -2.0$^{d)}$, -2.09$^{f)}$  & 8\arcmin.5$\pm$0\arcmin.5  & -7.9$\pm$0.5  \\
Canes Venatici II & 150$\pm$15  & 20.9$\pm$0.2  & -2.3$^{f)}$  & $\sim$3\arcmin.2  & -4.8$\pm$0.6  \\
Coma Berenices & 44$\pm$4 & 18.2$\pm$0.2 &  $\sim$-2, -2.00$^{f)}$  & $\sim$5\arcmin.5  & -3.7$\pm$0.6  \\
Hercules & 140$\pm$13  & 20.7$\pm$0.2  & -2.27$^{f)}$  & $\sim$8\arcmin.2  & -6.0$\pm$0.6  \\
Leo IV & 160$\pm$15  & 21.0$\pm$0.2  & -2.3$^{f)}$  & $\sim$3\arcmin.4  & -5.1$\pm$0.6  \\
Leo T & $\sim$420  & $\sim$23.1  & $\sim$-1.6, -2.3$^{f)}$  & $\sim$1\arcmin.4  & $\sim$-7.1  \\
Segue 1 & 23$\pm$2  & 16.8$\pm$0.2  & -  & $\sim$4\arcmin.5  & -3.0$\pm$0.6  \\
Ursa Major I & $\sim$100  & $\sim$20  &  $\sim$-2, -2.2$^{d)}$, -2.1$^{f)}$  & $\sim$7\arcmin.8  & $\sim$-6.5 \\
Ursa Major II & 32$\pm$5  & 17.5$\pm$0.3  & $\sim$-1.3, -1.6$^{d)}$, -2.0$^{f)}$  & $\sim$12\arcmin  & -3.8$\pm$0.6  \\
Willman 1 & 38$\pm$7  &  17.9$\pm$0.4 & $\sim$-1.3, -1.6$^{d)}$  & $\sim$1\arcmin.9  & $\sim$-2.5  \\
\tablecomments{Values are from the respective discovery papers (see
  text), unless otherwise indicated: a) \cite{dallora06}; b)
  \cite{siegel06}; c) \cite{munoz06}; d) \cite{martin07}; e) \cite{ibata06}; f) \cite{simon07}  }
\enddata
\label{tab:newprops}
\end{deluxetable*}

\subsection{Fitting methods and results}

\subsubsection{Single component fits}
\label{subsec:nsscfits}

\begin{figure*}
\epsscale{1.0}
\plotone{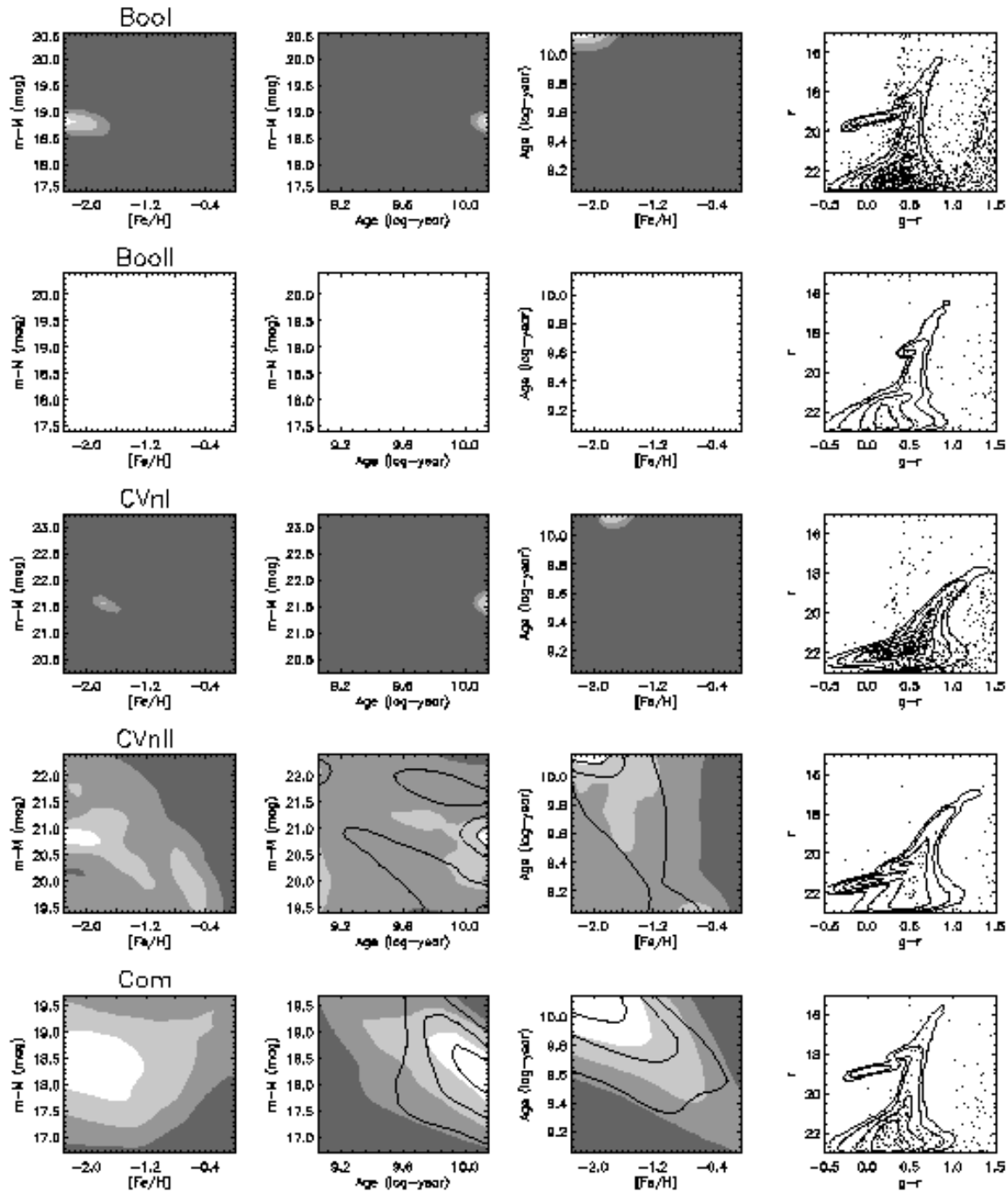}
\caption{Population properties of newly discovered Milky Way
  satellites: the panels show single component fitting results for Boo
  I, Boo II, CVn I, CVn II and Com, with the columns showing (from
  left to right): distance vs. metallicity; distance vs. age; age
  vs. metallicity; and the color-magnitude diagram with the best-fit
  single component model in contours. The gray-scale contours show the
  regions in parameter space that are within 1$\sigma$, within
  2$\sigma$, within 3$\sigma$ and outside 3$\sigma$ of the
  best-fitting single component model, going from white to dark
  gray. The solid line contours are for the case in which one of
  the parameters is fixed to the literature value (i.e. age for the
  left-most, metallicity for the middle-left, and distance for the
  middle-right columns), where available. }
\label{fig:nsscfit1}
\end{figure*}

\begin{figure*}
\epsscale{1.0}
\plotone{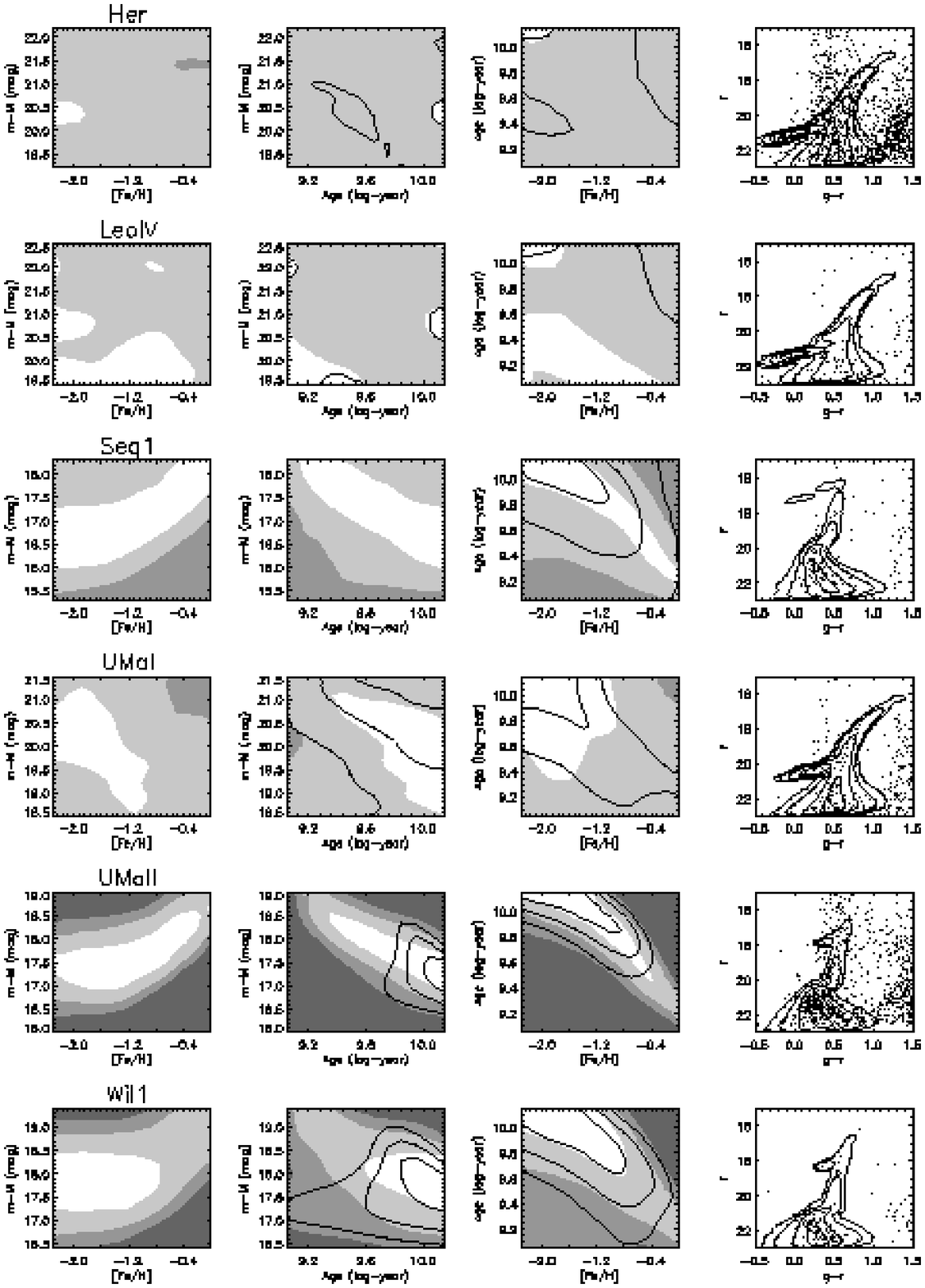}
\caption{Same as Figure \ref{fig:nsscfit1}, but for Her, Leo IV, Seg 1, UMa
  I, UMa II and Wil 1.
}
\label{fig:nsscfit2}
\end{figure*}

As for the GCs we obtained data in 5\degree$\times$5\degree~ fields around
each object from the SDSS CAS as outlined in section
\ref{sec:gctests}.  For the target CMDs we used the stars within one
half-light radius ($r_h$, Table \ref{tab:newprops}) from the center
(with a minimum of 3\arcmin), with all
stars outside a radius of 60\arcmin~ used for the control CMDs. Figure
\ref{fig:nscmds} shows the target CMDs for the twelve objects. In all
cases except Leo T a color range of $-0.5<g-r<1.5$ is used for the
CMD fits. The population parameter space we sample consists of 12 age bins, 24
metallicity bins and 31 distance bins. The age bins have a 0.1
log-year width and separations of 0.1 log-year, with the oldest
running from 10.1--10.2 log-years (12.5 Gyr to 15.8 Gyr) and the
youngest from 9.0--9.1 (1 Gyr to 1.26 Gyr). Metallicities run from
[Fe/H]=$-$2.3 to 0.0 in steps of 0.1 dex, and each metallicity bin
has a width of 0.2 dex. The range in distance modulus probed is 3
magnitudes wide, with a resolution of 0.1 magnitudes, and is centered on
the literature value for each object.

For Leo T we use a slightly different approach.  The group of
blue stars in the CMD is too bright and does not have the right
morphology to be interpreted as HB stars belonging to the same
population as the RGB stars. This means there are two populations
present in two different parts of the CMD, with the RGB corresponding
to the old population and the stars at $g-r \lesssim$0.3 to a much
younger population. As demonstrated in Section \ref{subsec:fakegc}, in
such a case the SC fits will give very poor results. However, since
the two populations in Leo T are only present on one side of the $g-r
\simeq 0.3$ line, they can be easily separated. Thus, to determine
the properties of the two populations separately we divide the Leo T
CMD into a red (-1.0$< g-r <$0.3) and a blue (0.3$< g-r <$1.5) part
and fit these separately. The red part of the CMD is fit with the same
parameter space as the other objects. For the blue part the same
metallicities and distance moduli are probed, but a younger set of age
bins is used. Here we use nine bins, the four oldest of which have a
width of 0.1 log-years centered on 9.05, 9.15, 9.25 and 9.35; the five
youngest have a width of 0.2 log-years centered on 8.1, 8.3, 8.5, 8.7
and 8.9 log-years.

\begin{figure*}[ht]
\epsscale{1.0}
\plotone{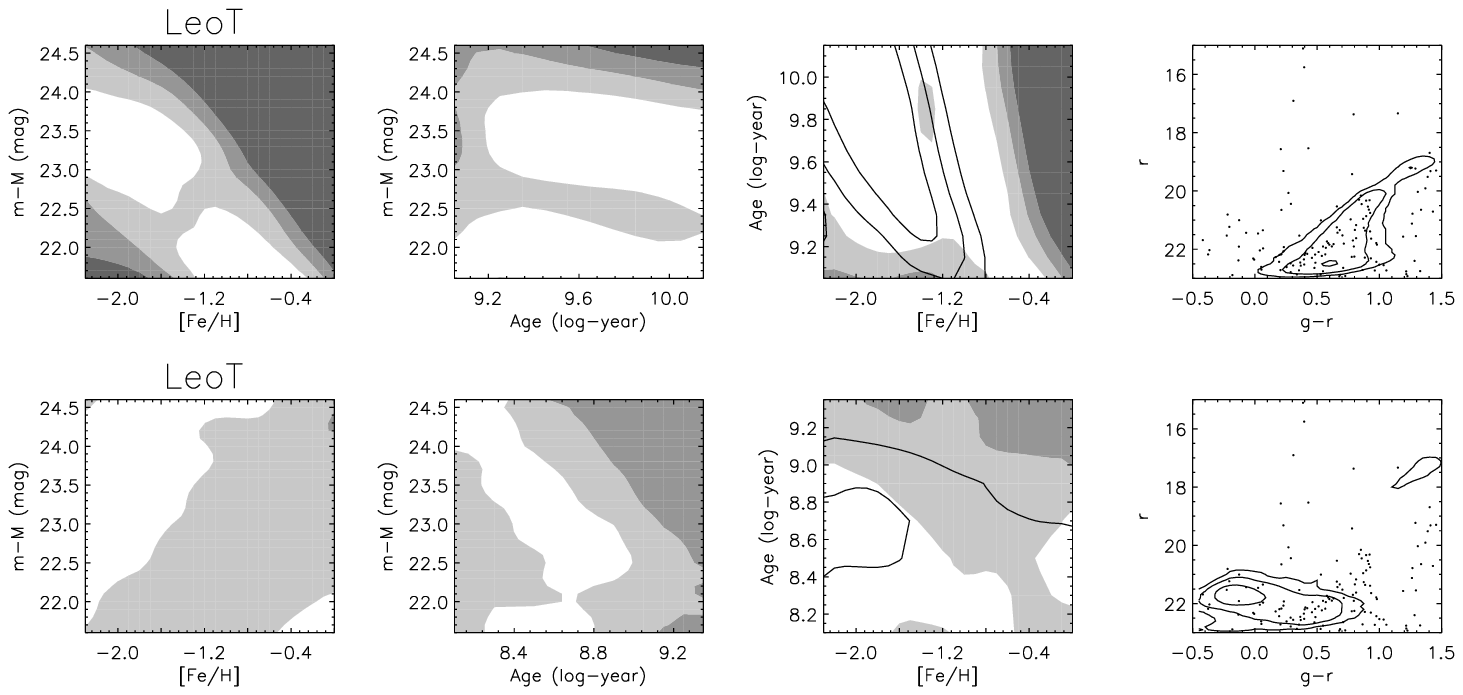}
\caption{Same as Figure \ref{fig:nsscfit1}, but for the red part
  ($g-r>0.3$) of the Leo T CMD (upper panels) and for the blue part
  (lower panels).
}
\label{fig:leoTscfit}
\end{figure*}

In Figures \ref{fig:nsscfit1} and \ref{fig:nsscfit2} the SC fitting
results for the new MW satellites, except Leo T, are presented. The
right-most panels show the CMD of each object again, with the best-fit
SC model overplotted as contours. Monte Carlo runs were used to
determine the values of $\sigma$. Except for Boo I and CVn I, solid
contours are shown for the case of a fixed metallicity or age, where
such a measurement is available in the literature.  Figure
\ref{fig:leoTscfit} shows the same results for the red and blue part
of the Leo T CMD.  The fact that low S/N hinders a robust
determination of the population properties provides a serious problem
for the majority of the targets. Although for most objects a finite
region of parameter space lies within 1 $\sigma$ from the best fit
(the white area in the contour plots),
reasonable measurements of any individual parameter can only be obtained
for Boo I, CVn I, CVn II, Com and Her. Especially for Boo II, Leo IV
and Leo T, parameters are very poorly constrained, and $Q$ space shows
a complicated morphology with multiple minima.

\subsubsection{Star formation histories}
\label{subsec:nssfh}

\begin{figure*}[ht]
\epsscale{1.0}
\plotone{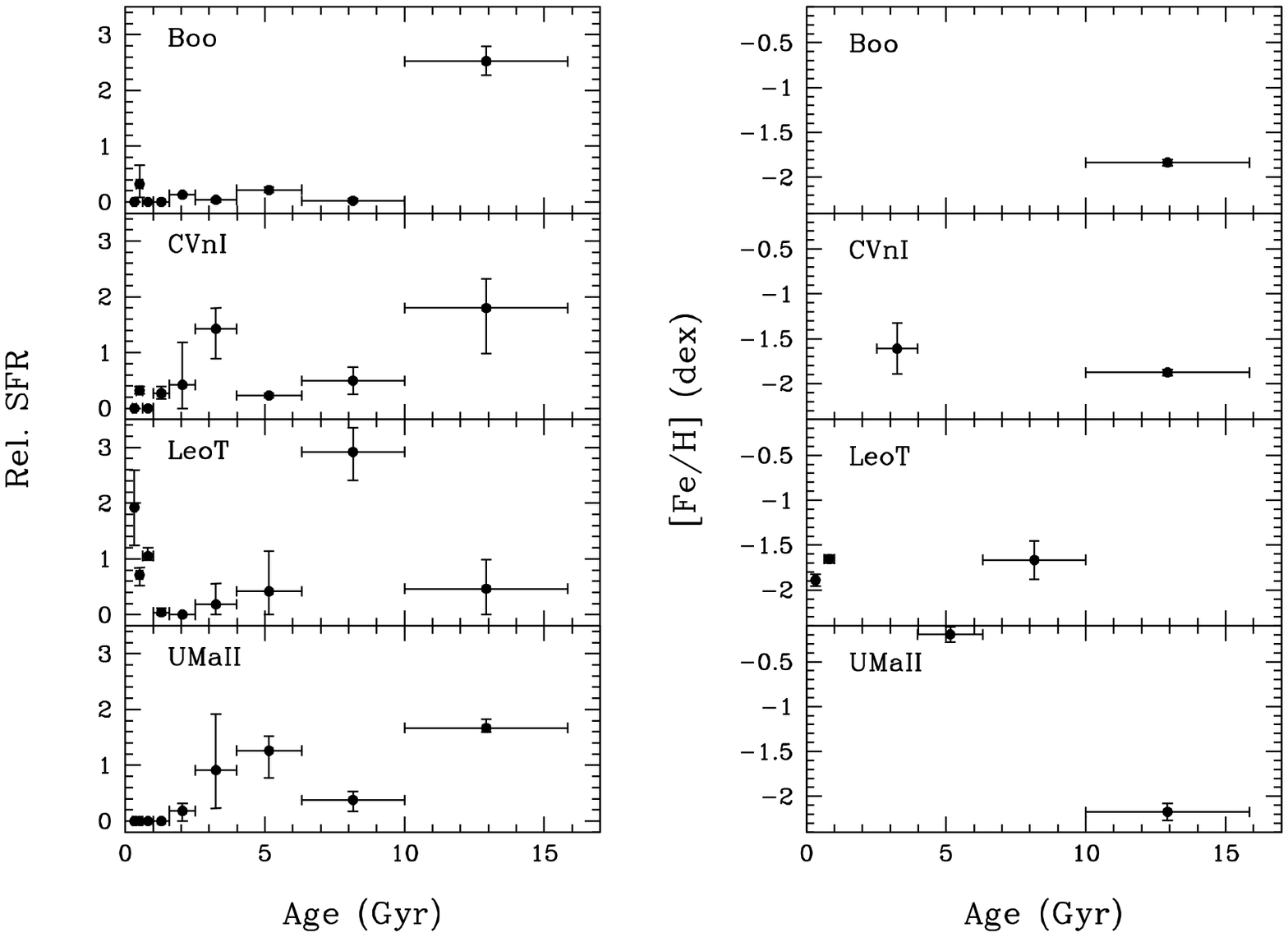}
\caption{Same as Figure \ref{fig:gcsfh} but showing the star formation histories and age metallicity relations for Boo, CVn I, Leo T and UMa II.
}
\label{fig:nssfh}
\end{figure*}

To constrain the SFHs of Boo I, CVn I, Leo T and UMa II we follow the
same approach as before for the GCs. A grid composed of nine age bins
and twelve metallicity bins is used, the age bins going from 10.2 to
8.4 log-years with an 0.2 log-year width, and the metallicities from
[Fe/H]=-2.4 to 0.0 with a 0.2 dex width. A fit is done at three
different distance moduli, 0.1 magnitude apart, with the middle value
equal to the literature value.  The resulting SFHs and AMRs are
plotted in Figure \ref{fig:nssfh}.

\subsubsection{Object detection}
\label{subsec:nsarea}

As with the GCs, we also test our overdensity detection scheme,
described in Section \ref{subsec:detect}, on these new Milky Way
satellites. The 5\degree$\times$5\degree~ field centered on each
object is divided in a square grid of 225 20\arcmin$\times$20\arcmin~
boxes, the CMDs of which are fit both with a control field and with
the control field plus a single component population model. As the control
field the whole 5\degree$\times$5\degree~ field is used.
For the SC model we consider two cases. First we use a typical old
population model, with age range 10.0 to 10.1 log-year and
[Fe/H]$=2.0$ at the best-fit distance obtained with the SC fits from
Section \ref{subsec:nsscfits}. Second we use a model corresponding to
the best-fit properties obtained for each object. 
The former case serves to simulate a realistic search of all SDSS data
for unknown stellar overdensities. When performing such a search, an
old, metal-poor model would be a likely choice since most MW
satellites are dominated by such populations. Another factor that
would influence the results of a search for unknown stellar
overdensities is the size of the box used. As a test, we also
did the same exercise using different box sizes for Boo I.
In addition to the 20\arcmin$\times$20\arcmin~ boxes,
we also used 40\arcmin$\times$40\arcmin~ boxes and 1\degree$\times$1\degree
boxes. To enable the use of these coarser grids, we retrieved a
larger, 10\degree$\times$10\degree~ field from the SDSS DR5 archive
centered on Boo I.
 
\begin{figure*}[ht]
\epsscale{1.0}
\plotone{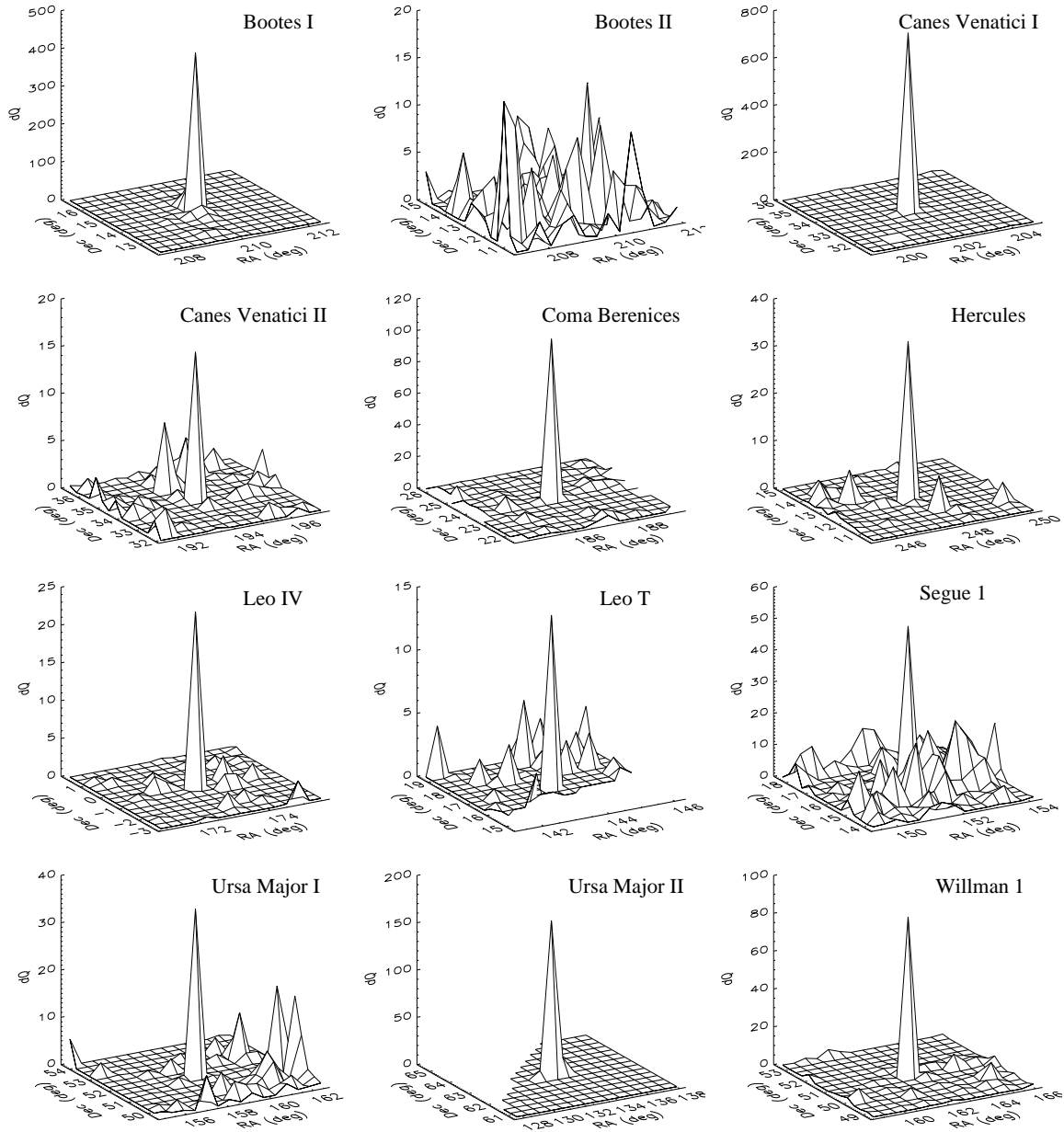}
\caption{Using MATCH to detect or verify stellar overdensities: Each
  panel shows (for one of the Milky Way satellites) the difference in
  fit quality, $dQ$, between fitting only a control field and fitting
  a control field plus SC model to the CMDs of a grid of boxes on the
  sky. For this, the 5\degree$\times$5\degree~ fields centered on each
  target were divided in 225 20\arcmin$\times$20\arcmin~ boxes. A high
  value of $dQ$ indicates a large improvement and the presence of a
  large number of stars that are well-fit by the SC model.  }
\label{fig:nsarea}
\end{figure*}

\begin{figure*}[ht]
\epsscale{1.0}
\plotone{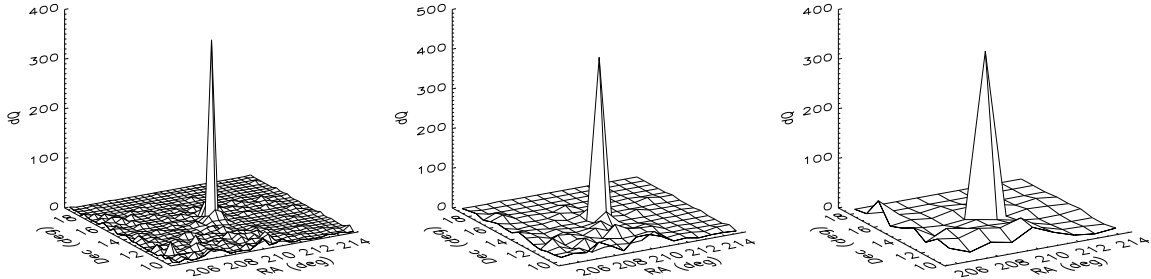}
\caption{Overdensity detection applied to Boo I using different
  spatial resolutions. A 10\degree$\times$10\degree~ field centered on
  Boo I is divided into 20\arcmin$\times$20\arcmin~ (left panel),
  40\arcmin$\times$40\arcmin~ (middle panel), and
  1\degree$\times$1\degree~ (right panel) boxes. The fit improvement,
  $dQ$, between a pure control field fit and a control field plus SC
  model fit, is plotted as function of position on the sky.
}
\label{fig:BooIareas}
\end{figure*}

In Figure \ref{fig:nsarea} we present the results of the fits with the
fine-tuned stellar population models to the twelve objects. Each panel
shows the improvement in the goodness-of-fit, $dQ$, as function of the
position in the field centered on a satellite.  Figure
\ref{fig:BooIareas} shows the resulting $dQ$ for the three different
box sizes in a 10\degree$\times$10\degree~ field centered on Boo I.
The significances of all detections, expressed in $dQ/rms$,
are listed in Table \ref{tab:detsig}.

\begin{deluxetable}{lcc}
\tablecaption{Detection significance for new Milky Way satellites}
\tablewidth{0pt} 
\tablehead{ \colhead{Object} & \colhead{Old $dQ/rms$} &
  \colhead{Fine-tuned $dQ/rms$} }
\startdata
Boo I$^{a)}$ & 106.9 & 121.6, 58.2, 27.8 \\
Boo II & 0.86 & 0.93 \\
CVn I & 447.7 & 589.4 \\
CVn II & 2.61 & 13.3 \\
Com & 57.5 & 53.9 \\
Her & 15.2 & 28.2 \\
Leo IV & 13.8 & 38.2 \\
Leo T & 5.8 & 11.3 \\
Seg 1 & 9.9 & 7.8 \\
UMa I & 12.9 & 11.2 \\
UMa II & 1156.2 & 865.3 \\
Wil 1 & 48.8 & 47.8 \\
\tablecomments{a) the three values of $dQ/rms$ for the fine-tuned
  model fits to Boo I correspond to the
box sizes of 20\arcmin$\times$20\arcmin~, 40\arcmin$\times$40\arcmin~
and 1\degree$\times$1\degree, respectively.}
\enddata
\label{tab:detsig}
\end{deluxetable}

\subsection{Discussion}

\subsubsection{Satellite properties}
\label{subsec:nsprops}

We have performed a systematic and uniform numerical CMD analysis of
the recently discovered MW satellites. How well the properties of an
individual object can be determined depends on the number of member
stars present in the CMD and the observed CMD features. 
For some objects, this type of analysis is
hard-pressed to provide strong statistical constraints due to the
sparseness of their CMDs. In Table \ref{tab:newresults} we summarize
the results of our single component and star formation history
fits. The distance moduli, metallicities and ages listed are the
unweighted averages of all SC fits that lie within 1 $\sigma$ of the
best SC fit and are therefore considered to be acceptable fits. The
errors are the standard deviations of the parameter values of these
fits, combined with an added systematic uncertainty of 0.1 dex in
[Fe/H] and 0.1 mag in distance modulus to account for isochrone
uncertainties. In cases where these properties cannot be sufficiently
constrained we list the metallicities and ages we get when assuming
the distance modulus from the literature, as this is usually the
best-determined property. The star formation history results, where
available, are summarized in Table \ref{tab:newresults} by the main
star formation episodes. Below follows an object-by-object discussion
of these and the star formation history results.

{\it Boo I.}--- This dwarf spheroidal is relatively nearby at $\sim$60
kpc and an RGB, HB and MSTO are all present in the SDSS data,
albeit sparsely sampled. The overall properties can therefore be
determined with relatively good accuracy, and the star formation
history is the best-constrained of all.  Especially the distance
of Boo I is very well constrained thanks to the HB which is detected
at high S/N due to the absence of field stars at $g-r <$0. The MSTO
and slope of the RGB provide constraints on the age and metallicity,
respectively.  Our distance, metallicity and age measurements agree
very well with the findings of other authors based on spectroscopy and
deeper imaging data. The star formation history shows one significant
epoch of star formation in the oldest ($>10$ Gyr) bin, with no
other bins containing a significant SFR. Considering the ability of
these fits to distinguish between a single and a multiple component
system, demonstrated by the fits to the artificial CMDs in Section
\ref{subsec:fakegc}, we conclude, despite the sparseness of the CMD,
that Boo I very probably only contains old stars.  Thus, our results
are consistent with the picture of a purely old, metal-poor stellar
population.

{\it Boo II.}--- The number of Boo II member stars in the SDSS data is
very small and traces only the RGB. Unfortunately, the resulting very
low S/N combined with the degeneracies between the different
parameters make it impossible for MATCH to constrain the stellar
population parameters in a statistically significant way. Although
\cite{BooII} show that the SDSS CMD of Boo II is {\it consistent} with Boo
II having similar metallicity, age, and distance as Boo I, The SDSS
data alone do not support any significant constraints on any of these
properties. The only way to get such constraints is by obtaining
deeper follow-up data.

{\it CVn I.}--- Being the brightest of the twelve objects analyzed
here, CVn I has a well-populated RGB and HB. This helps to
increase the robustness of the constraints we obtain from the SC
fits.  Although at the edge of the sensitivity limits of SDSS,
the HB enables the determination of the distance. Moreover, the extent
of the HB to very blue colors requires the stars in CVn I to be both
old and metal-poor. This constraint on the metallicity is further
enhanced by the slope of the RGB. Together, the extended HB and RGB
thus can provide robust constraints on the properties of the dominant
stellar population.  The distance modulus we find is slightly smaller
than that found in the literature, but not significantly so. The
metallicity of [Fe/H]=-1.8$\pm$0.2 falls nicely within the range of
metallicities observed by \cite{ibata06}.  Although an age of 14$\pm$2
Gyr fits the CMD best, the star formation history shows two peaks.
The oldest star forming episode in the 10-16 Gyr bin is dominant in
terms of the number of stars and drives the SC fits, but the peak in
the 2.5-4 Gyr bin is equally significant. For the old stars the
metallicity is found to be close to [Fe/H]=-2, while the younger stars
have metallicities around [Fe/H]=-1.5.  It should be noted that
the young population found here is not caused by the algorithm
attempting to fit BHB stars, as is the case for M15 in Figure
\ref{fig:gcsfh}. Stars belonging to this young population only appear
in the regions of the CMD where the RGB and the red part of the HB are
located. The presence of two populations is therefore purely based on
the RGB morphology and the density of stars along the HB.  This
detection of an old and intermediate age population with these
metallicities confirms the findings of \cite{ibata06} and
\cite{cvnIlbt}. Furthermore,
\cite{CVnI} note the presence of a carbon star very close to the
center of CVn I and its heliocentric radial velocity of 36$\pm$20 km/s
is consistent with the radial velocities that \cite{ibata06} and
\cite{simon07} measure for CVn I, making it is likely that this star
belongs to the CVn I dwarf. Since carbon stars are often of
intermediate age, this is another indication of the presence of an
intermediate age population in this system.

{\it CVn II.}--- Although very faint and distant, our distance
measurement for CVn II is very accurate and in very good agreement
with the value from \cite{5pack} due to the presence of HB stars.
From the extension of the HB to very blue colors ($g-r \sim -0.3$) we
can infer that the stellar population is both old and
metal-poor. Despite the sparseness of the CMD we thus get relatively
good constraints on these parameters and we find a metallicity of
[Fe/H]$\sim$-2 and a very high age. CVn II thus appears to be a very
old and metal-poor object.

{\it Com.}--- An absence of a clear HB makes constraining the distance
of Com difficult and leads to a strong degeneracy between age,
distance and metallicity. Similar degeneracies occur for
artificial CMD 1 in Section \ref{subsec:fakegc} and the GC NGC 6229
(see Figures \ref{fig:fakescfit} and \ref{fig:gcsfh}), even though
their CMDs are much less sparsely populated. The RGB and MSTO alone do
not allow a robust determination of the individual population
parameters. As a consequence, the distance modulus we recover has
rather large error bars but is consistent with the findings of
\cite{5pack}. The SC fits favor an ancient population ($>10$ Gyr)
and a metallicity of [Fe/H]$\sim -$2, but also here the uncertainties
are large.

{\it Her.}--- In the case of Her the contamination of the CMD with
fore- and background stars is severe, which is due to its location at
lower Galactic latitudes compared to the other objects. This high
contamination means a low S/N of the CMD features of Her, apart from
the HB, which is in the relatively clean part of the CMD at $g-r <
0$. Our distance modulus of 20.4$\pm$0.2, which largely is
determined by the HB, is somewhat lower than estimated by
\cite{5pack}, but is consistent within the error bars. This object too
is consistent with an ancient and metal-poor ([Fe/H]$\sim$-2)
stellar population. Even though the region of parameter space
that lies within 1 $\sigma$ of the best fit values is well
constrained, the complete parameter space lies within 2 $\sigma$,
which is due to the low S/N of the CMD. These results based only on
SDSS data are also consistent with the results obtained by applying
the same methods to much deeper $B,V$ photometry from the Large
Binocular Telescope \citep{coleman07}. In fact, the uncertainties
on the values of distance, metallicity and age from the deeper data
are very similar to the 1$\sigma$ values we obtain here.

\begin{deluxetable*}{lcccr}
\tablecaption{New MW satellite results}
\tablewidth{0pt} 
\tablehead{ \colhead{Object} & \colhead{m-M} & \colhead{[Fe/H]} &
  \colhead{Age} & \colhead{SF episodes} }
\startdata
Bootes I (BooI) & 18.8$\pm$0.2 & -2.2$\pm$0.2 & 14$\pm$2 & $\sim$10-16 Gyr\\
Bootes II (BooII) & - & - & - & - \\
Canes Venatici I (CVnI) & 21.5$\pm$0.2 & -1.8$\pm$0.2 & 14$\pm$2 &
$\sim$2-4 Gyr, $\sim$10-16 Gyr\\
Canes Venatici II (CVnII) & 20.8$\pm$0.2 & -2.1$\pm$0.3 & 14$\pm$2 & - \\
Coma Berenices (Com) & 18.4$\pm$0.4 & -1.9$\pm0.4$ & 11$\pm5$ & - \\
Hercules (Her) & 20.4$\pm$0.2 & -2.1$\pm$0.2 & 14$\pm$3 & - \\
Leo IV (LeoIV) & - & {\it -2.1$\pm$0.3} & {\it 14$\pm$2} & - \\
Leo T (LeoT) & 23.1$\pm$0.6 & -1.6$\pm$0.6 & - & $<1$ Gyr, $\sim$6-10 Gyr \\
Segue 1 (Seg1) & - & {\it -1.6$\pm$0.5} & {\it 11$\pm$4} & - \\
Ursa Major I (UMaI) & - & {\it -1.8$\pm$0.4} & {\it 10$\pm$5} & - \\
Ursa Major II (UMaII) & - & {\it -1.5$\pm$0.5} & {\it 11$\pm$4} & $\sim$2-6 Gyr, $\sim$10-16 Gyr \\
Willman 1 (Wil1) & 17.9$\pm$0.4 & -1.6$\pm$0.5 & 10$\pm$5 & - \\
\enddata
\label{tab:newresults}
\end{deluxetable*}

{\it Leo IV.}--- The CMD of Leo IV is so sparse that a
statistically significant determination of the population properties
is not possible based on the SC fits alone. The best fit is given by a
population of $\sim$14 Gyr old stars with [Fe/H]=$-$2.2 at a
distance modulus of 20.8, but a large number of models gives a
goodness-of-fit that is only marginally worse. Assuming a distance
modulus of 21.0 \citep{5pack}, gives the solid contours in the
age-metallicity panel for Leo IV in Figure \ref{fig:nsscfit2}. Using
this prior, the SC fits allow the age and metallicity to be
constrained to 14$\pm$2 Gyr and [Fe/H]$=-$2.1 $\pm$0.3,
respectively.

{\it Leo T.}--- This object is very interesting and differs from the
others analyzed here, in that it is the least luminous system known
which shows signs of on-going star formation and has HI gas associated
with it \citep{LeoT}. Because of the presence of two clearly very
different populations the SC fitting method is not very well
suited. Even when fitting the two populations separately, by
dividing the CMD in two distinct parts as we have done, the results
are poor, due to the small numbers of stars in each part. Taking the
results of the two populations together we find a distance modulus of
23.1$\pm$0.6 magnitudes. Both the old and young populations are best
fit with low metallicities, [Fe/H]$\lesssim$-1.5. The star formation
history we find (Figure \ref{fig:nssfh}) clearly shows two peaks,
confirming the presence of two distinct populations. The age of the
old population, not well-constrained with the SC fits, is put in the
6.5-10 Gyr age bin. Comparing this with the SFHs of the GCs in
Figure \ref{fig:gcsfh} and the other SFHs in Figure \ref{fig:nssfh},
where in all cases the peak of star formation is in the oldest age
bin, this implies that the old population in Leo T is on average
younger than the old population in these other objects. The young
stars are clearly found in the three youngest age bins, where they are
fit as young ($<1$ Gyr) blue loop stars. \cite{LeoT} interpreted the
nature of these stars similarly.

{\it Seg1.}--- In Seg 1 only a MS and MSTO are populated in the
very sparse CMD. As with the artificial CMD 1 in Section
\ref{subsec:fakegc} and the GCs M13, M15 and M92, where the SC fits
also solely relie on the MS and MSTO, strong degeneracies between age,
distance and metallicity are expected. Indeed, the contour plots for
Seg 1 in Figure \ref{fig:nsscfit2} show a similar morphology, but due
to the sparseness of the CMD with even poorer constraints. This makes
any determination of the population properties of Seg 1, based solely on
the SC fits, impossible. Even when using the distance modulus from
\cite{5pack} as a prior, the age-metallicity degeneracy prevents
strong constraints on these parameters, resulting in very large error
bars on the values for Seg 1 listed in Table \ref{tab:newresults}.

{\it UMa I.}--- This object also has a very sparsely populated
CMD. An RGB is readily seen, with some probable HB stars that,
however, are too few to produce strong statistical constraints on the
distance. Because of this and the usual degeneracies, the
significance of the SC fit results is very low, causing all parameters
to be poorly constrained. Assuming a distance modulus of 20
\citep{UMaI}, our results are consistent with an old ($\sim$10 Gyr)
and metal-poor population, but with large error bars.

{\it UMa II.}--- Although the MSTO of UMa II is well-populated,
the scarcity of stars in other parts of the CMD provides problems for
this object. There are some RC stars, but the S/N of this CMD feature
is not high enough to make for a reliable distance
determination. Added to that there are strong degeneracies between the
different solutions; similar degeneracies are seen in Section
\ref{sec:gctests} for the GC NGC 6229 where also the MSTO drives the SC fit
results. Our results are, however, consistent with previous
findings. With a prior on the distance modulus, $m-M=$17.5, the fits
agree with the conclusion of \cite{UMaII} that UMa II is not very
metal-poor, but seems to have a more intermediate metallicity of
[Fe/H]$\sim-$1.5, although the error bars are large (see Table
\ref{tab:newresults}). The spectroscopic metallicity measurement from
\cite{martin07} is consistent with our value, but the one from
\cite{simon07} is more metal-poor, but still consistent within
the errors. While the SC fits imply that the majority of stars
is likely old, $\sim$11 Gyr, the SFH (Figure \ref{fig:nssfh}) hints at
more complexity. 
According to the SFH results, there is a population of very old
($>$10 Gyr) and metal-poor ([Fe/H]$\simeq$-2), as well as a younger
($\sim$5 Gyr), close to solar metallicity population. Since the SC fits
measure the average properties, this is consistent with the
intermediate metallicity we find there. \cite{UMaII} report that the
upper main sequence of UMa II in their deeper Subaru data is wider
than expected just from observational errors, one explanation for
which could be the presence of a mix of populations. 
Also in our fit, the double peak results from the width of the
MSTO, which is less well-fit by a single-age population. The exact
ages and metallicities of the populations should be studied with
deeper and spectroscopic data, since the SDSS CMD does not support
such detailed conclusions. However, taking our SFH fit and these
other results together it does seem likely that UMa II, like CVn I, is a
system with a complex star formation history, which might also be (part
of) the explanation for the discrepancy between the two available
spectroscopic metallicity measurements.

{\it Wil1.}--- For this system only the MSTO is observed, which
is also sparsely populated. This leads to a strongly degenerate
solution space, similar to the cases of UMa II and Seg 1.
The distance can be determined slightly better here, but the range of
distance moduli for which fits within 1$\sigma$ from the best fit can
be obtained is still very large: 17.2$<m-M<$18.5. Using the
distance modulus from \cite{Wil1} (m-M=17.9) as a prior, the fit
results imply an old population ($\sim$10 Gyr) with an intermediate to
low metallicity ([Fe/H]$\sim$-1.6), but also here the error bars
on these numbers are large.

\begin{figure*}
\epsscale{1.0}
\plotone{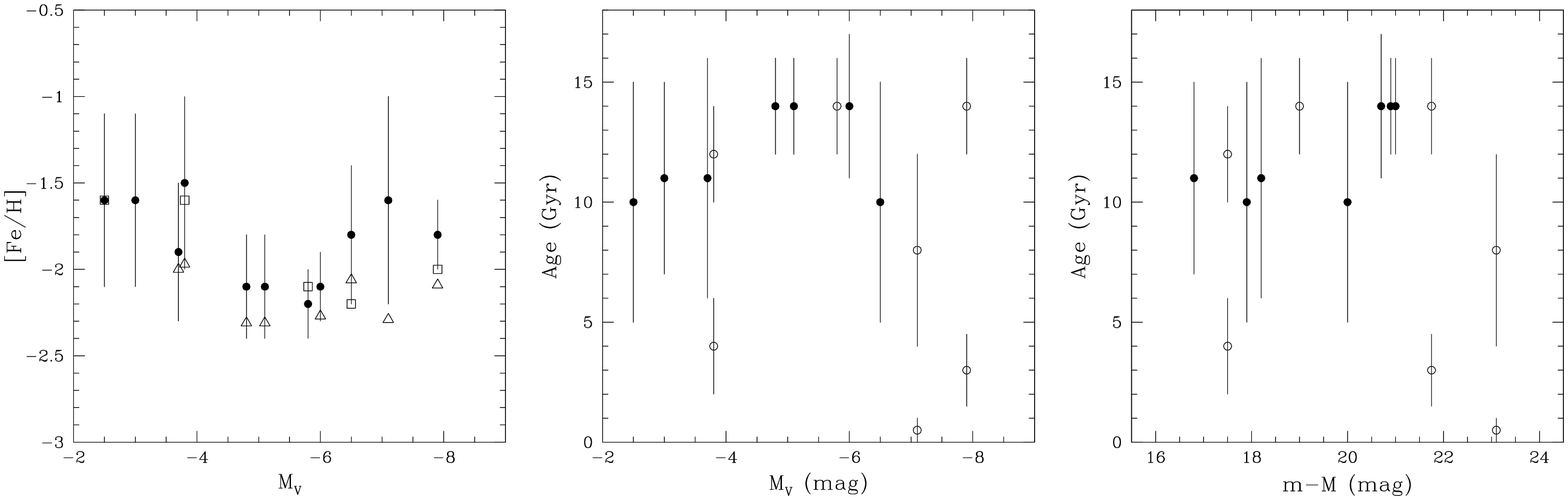}
\caption{CMD fitting results for new Milky Way satellites: {\it Left:}
  metallicities from this paper (filled circles) versus absolute
  magnitude, compared with spectroscopic metallicities from
  \cite{martin07} (open squares) and \cite{simon07} (open
  triangles). For clarity, only the error bars on the results from
  this work are shown. {\it Middle:} ages of the satellites versus
  absolute magnitude. Open circles denote objects for which a full SFH
  fit was done, and in the cases where two distinct populations were
  found (CVn I, Leo T and UMa II) both are plotted; filled circles
  denote objects where the age is only determined from the SC fits.
  {\it Right:} as in the middle panel, the ages are plotted, but now
  versus distance modulus.
}
\label{fig:nssummary}
\end{figure*}

Figure \ref{fig:nssummary} summarizes our results from the MATCH fits
to the CMDs for the sample of new Milky Way satellites.  In the left
panel we plot our metallicity estimates for all objects versus their
absolute magnitudes, together with the spectroscopic metallicities
from \cite{martin07} and \cite{simon07}. Especially for the
fainter objects our metallicity estimates are very poorly constrained
(Figures \ref{fig:nsscfit1} and \ref{fig:nsscfit2}) and because of
the available isochrones our metallicity range is restricted to
[Fe/H]$=-$2.4 at the metal-poor end. In light of this, our photometric
results agree remarkably well with the spectroscopic values. Still,
care has to be taken when drawing conclusions from these metallicity
estimates. We can say at best that we find no indication for a
corellation between metallicity and the luminosity of the objects,
which is observed for brighter dwarf spheroidals
\citep[e.g.,][]{grebel99,grebel03}.  The mean age of the stars versus
absolute magnitude is shown in the middle panel of Figure
\ref{fig:nssummary}. For objects in which the SFH fits showed evidence
for two populations, both populations are plotted separately.  The
right panel shows the inferred ages of the satellites versus distance
modulus, to probe possible environmental influences.

Looking at the middle panel of Figure \ref{fig:nssummary} there is no
clear evidence for a luminosity dependence of the ability of a dwarf
galaxy to form stars recently. Although the two most luminous
objects in this sample, Leo T and CVn I, both have young or
intermediate age populations, the much fainter UMa II does as well.
While the environment might play a role, the right panel of Figure
\ref{fig:nssummary} shows no obvious relation between age or the
presence of multiple star formation epochs with distance from the sun
(since these objects are at high Galactic latitude and at large
distances, the heliocentric and galactocentric distances are not
significantly different). This is contrary to the trend suggested by
\cite{vandenbergh94} for the brighter Local Group dwarf spheroidals,
in which systems with a larger fraction of young or intermediate age
stars are systematically more distant.

\subsubsection{Satellite detection}
\label{subsec:nsdetect}

\begin{figure}
\epsscale{1.0}
\plotone{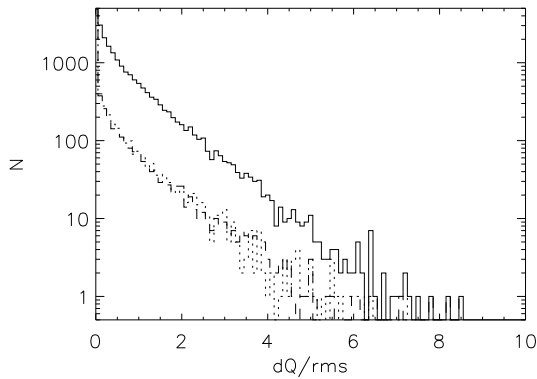}
\caption{Histogram of the $dQ/rms$ distribution for an empty region of
  sky of effective area of $\sim$800 square degrees.
}
\label{fig:pksig}
\end{figure}

Since most of these recently detected objects are intrinsically faint
and/or distant, they are of course much harder to detect by any
overdensity detection scheme than the GCs.  The improvement in the
goodness-of-fit, $dQ$, between a fit with a control field only and a
control field plus SC model is roughly proportional to the number of
stars in the CMD that can be fit by the SC model. This means that the
objects with sparsely populated CMDs are the most difficult to
detect. Comparing the overdensity detection results in Figure
\ref{fig:nsarea} with the CMDs of the objects (Figure
\ref{fig:nscmds}) confirms this, as Boo, CVn I and UMa II give by far
the highest $dQ$. 
Similarly, the rms noise in $dQ$ increases with increasing numbers of
field stars in the CMDs, and at the low S/N end the noise level starts
to play an important role. For example, comparing the panels for Boo
II and CVn II in Figure \ref{fig:nsarea}, which have the same $dQ$
scale, reveals very different noise levels. This can be traced back
to the positions of these objects on the sky, with Bootes II lying
behind the Sagittarius stream and CVn II in a much `cleaner' region of
the halo. The Sagittarius stream also produces a high noise level
around Segue 1.

While most objects produce clear peaks in $dQ$, the detectability of
several is rather low. For example, CVn II, Leo T, Segue 1 and UMa I
have a $dQ$ only a factor of a few higher than the most prominent
noise peaks; Boo II is not clearly detected at all. When inspecting
the values in Table \ref{tab:detsig}, it turns out that the use of a
fine-tuned population model as compared to the generic old, metal-poor
model has a varied effect. In some cases the detection significance is
improved, in some cases the change has little effect and in a few cases the
detectability is actually decreased. In these last cases, even though
the peak $dQ$ is the same to or higher for the fine-tuned model, the
change in the rms causes the detection to become less significant.
This indicates that, while sampling a range of population models would be
preferable when doing an overdensity search, a relatively small grid
of models will not severely compromise the outcome.

To test the significance of the $dQ/rms$ measure, we tested the
overdensity detection scheme on a presumably `empty' patch of SDSS sky
with 170\degree$<$RA$<$200\degree and
40\degree$<$Dec$<$65\degree. This patch of sky was divided into
20\arcmin$\times$20\arcmin~ boxes with some overlap, giving a total of
7088 boxes, equalling an effective search area of almost 800 square
degrees. A control field (the CMD from the surounding
5\degree$\times$5\degree area) was fit to the CMD of each box, as well
as this control field plus a typical old stellar population model with
an age bin of 10.0 to 10.1 log-year and [Fe/H]=-2. This was done seven
times, with the stellar population model shifted to different distance
moduli: 17.0, 18.0, 19.0, 20.0, 21.0, 22.0 and 23.0 magnitudes. Thus, for
each box seven values of $dQ/rms$ are obtained, where the rms is
obtained from the boxes in the surounding
5\degree$\times$5\degree. The methodology closely resembles the
approach used for the MW satellites. Figure \ref{fig:pksig} shows the
histogram of all $dQ/rms$ values for all boxes (solid line), as well
as for two individual distance moduli, 18.0 (dotted) and 22.0 (dashed)
mag. All three histograms show the same distribution, implying that the
noise behaves similarly for all distance moduli. The tail of the
distribution extends to $dQ/rms \sim 8$. An inspection of the
locations with values higher than 7 reveals no obvious overdensities,
implying that this tail is indeed caused by noise. In order to obtain
a clean sample of stellar overdensity detections, a $dQ/rms$ threshold
of $\sim$9 seems appropriate. This means that when using a
sufficiently sampled grid of population models and distances, all new
MW satellites would be detected, with the exception of Boo II
(cf. Table \ref{tab:detsig}).

The results of the fits to Boo I with different box sizes, listed in
Table \ref{tab:detsig} and shown in Figure \ref{fig:BooIareas},
illustrate the effect of the spatial resolution of the CMD grid. Here,
the detectability is also strongly affected by the rms of the
background noise. Since the $dQ$ due to random variations in the field
population increases with increasing numbers of field stars, the rms
is higher for larger box sizes, favoring the smallest boxes used
here. The fact that the $\Delta Q/rms$ decreases roughly by a factor
of 2 as we move to larger box sizes is caused by the number of
background stars increasing by a factor 4, while the number of Boo I
stars in the central box does not change.  It is also interesting to
note that here the noise is dominated by the Sagittarius stream, which
runs accross the lower left corner of the panels in Figure
\ref{fig:BooIareas}.  The significance of the detections in Table
\ref{tab:detsig} can be boosted by decreasing the box sizes,
especially for the objects with the smallest (projected) size.

\section{Summary and conclusions}
\label{sec:summary}

We have shown that application of MATCH \citep{match} to Hess diagrams
or CMDs of SDSS data successfully recovers the known stellar
population properties of globular clusters (Section
\ref{sec:gctests}), within the limitations of the data. This gives confidence that despite inaccuracies in
the theoretical isochrones in isolated parts of the CMD, the overall
CMD-fitting technique still provides good and trustworthy
results. Apart from the standard star formation history fitting
method, we have also demonstrated three other ways of applying MATCH
to SDSS data:\\
1) Identifying the best fitting `single component' stellar population,
a mode useful for CMDs of very faint or distant objects with limited
S/N in their CMD;\\
2) Fitting the line-of-sight distribution of stars -- to keep the fit's
degrees of freedom constant, we fix the metallicity at each age bin; and\\
3) Algorithmic detection of objects, i.e. of a localized stellar
population overdensity in a particular (unknown) direction and at a
particular (unknown) distance. This is done by comparing the
difference in the goodness-of-fit between a CMD-fit to a control field
and a CMD-fit to a control field plus model population.

These tests show that the application of CMD-fitting
techniques to wide-field survey data is a very promising endeavor in
the light of SDSS and other surveys that will start yielding data
over the next few years. Further improvements in software and
theoretical isochrones can only increase the use and accuracy of such
work. 
On the basis of these tests and the application to known globular
clusters, we have applied our single component fitting technique to
twelve recently discovered Milky Way satellites.
The sparseness of several of the CMDs of these objects inhibits a
robust determination of the population parameters and the best fit
solutions are often not unique. In order to obtain good constraints
at least some CMD features such as the MSTO, HB and RGB should be
observed with sufficiently high S/N. For the least luminous objects
this is inherently problematic because of their lack of (evolved) stars.
For only four of the satellites, the CMDs had sufficiently high S/N to
warrant a star formation history fit.

The results of the population fits to the new Milky Way satellites,
summarized in Table \ref{tab:newresults}, are consistent with the
individual discovery and follow-up studies. When looking at the sample
as a whole, we find that the majority of these objects are
likely dominated by old ($\gtrsim$ 10 Gyr) and metal-poor
([Fe/H]$\sim$-2) stellar populations. For Boo, CVn II, Com, and Her
the single component fits converge to this picture, and for Boo the
full fit of a star formation history also confirms this. Seg 1,
UMa I, UMa II, and Wil 1 are not well constrained by the SC fits, but
using the distances from the literature as priors, they are also
consistent with similar ages and metallicities, although slightly
higher metallicities ([Fe/H]$\sim -$1.6) cannot be excluded.
The single component fits show CVn I to be old as well, but not
extremely metal-poor. However, the SFH reveals the presence of both an
old, metal-poor, as well as a younger ($\lesssim$3 Gyr) and more
metal-rich population. This is consistent with the findings of
\cite{ibata06}, who detect two kinematically distinct populations in
CVn I which differ in age and metallicity, and the detection of a
young component by \cite{cvnIlbt} in deeper photometry. The SFH of UMa
II also shows a peak at an age of $\sim$5 Gyr, which supports
suggestion of multiple populations in this object by \cite{UMaII},
who find the MSTO width of UMa II to be too large to be explained by
photometric errors alone.  Finally, in Leo T the presence of two
distinct populations is readily seen in the CMD and reproduced in the
star formation history. The RGB seen in this system belongs most
likely to an intermediate age ($\sim$5-10 Gyr) population with a
metallicity of [Fe/H]$\sim$-1.5 to -2, while the blue stars in the CMD
are most likely very young ($<$1 Gyr) blue loop stars with similar
metallicities. The objects studied here are all intrinsically
faint and sparse, and in some cases the population parameters can be
constrained only poorly. However, their star formation histories do
not appear uniform, nor do all of them appear to be exclusively
old and metal poor.

\acknowledgments

We thank Matthew Coleman for helpful discussions and the
anonymous referee for his useful comments that helped to clarify
this paper.
J.T.A.d.J. acknowledges support from DFG Priority Program 1177.
D.B.Z. was supported by a PPARC-funded rolling grant position.
E.F.B. thanks the Deutsche Forschungsgemeinschaft for their support
through the Emmy Noether Program.
V.B. was supported by a PPARC Fellowship.

    Funding for the SDSS and SDSS-II has been provided by the Alfred
    P. Sloan Foundation, the Participating Institutions, the National
    Science Foundation, the U.S. Department of Energy, the National
    Aeronautics and Space Administration, the Japanese Monbukagakusho,
    the Max Planck Society, and the Higher Education Funding Council
    for England. The SDSS Web Site is http://www.sdss.org/.

    The SDSS is managed by the Astrophysical Research Consortium for
    the Participating Institutions. The Participating Institutions are
    the American Museum of Natural History, Astrophysical Institute
    Potsdam, University of Basel, University of Cambridge, Case
    Western Reserve University, University of Chicago, Drexel
    University, Fermilab, the Institute for Advanced Study, the Japan
    Participation Group, Johns Hopkins University, the Joint Institute
    for Nuclear Astrophysics, the Kavli Institute for Particle
    Astrophysics and Cosmology, the Korean Scientist Group, the
    Chinese Academy of Sciences (LAMOST), Los Alamos National
    Laboratory, the Max-Planck-Institute for Astronomy (MPIA), the
    Max-Planck-Institute for Astrophysics (MPA), New Mexico State
    University, Ohio State University, University of Pittsburgh,
    University of Portsmouth, Princeton University, the United States
    Naval Observatory, and the University of Washington.

\end{document}